\documentclass[prb,superscriptaddress,longbibliography,twocolumn]{revtex4-1}
\usepackage{geometry}
\usepackage{amsmath}
\usepackage{amssymb}
\usepackage{graphicx}
\usepackage{wasysym}
\usepackage{braket}
\usepackage{bm}
\usepackage[usenames,dvipsnames]{color}
\usepackage[caption=false,singlelinecheck=false]{subfig}
\usepackage[breaklinks,colorlinks = true,linkcolor = red,urlcolor=cyan,citecolor=red]{hyperref}

\graphicspath{./}

\newcommand{\be}{\begin{equation}}
\newcommand{\ee}{\end{equation}}

\newcommand{\bea}{\begin{eqnarray}}
\newcommand{\eea}{\end{eqnarray}}
\newcommand{\beal}{\begin{align}}
\newcommand{\eeal}{\end{align}}
\newcommand{\ra}{\rangle}
\newcommand{\la}{\langle}

\newcommand{\upa}{\uparrow}
\newcommand{\dna}{\downarrow}

\newcommand{\pdg}{{\phantom\dagger}}

\newcommand{\fQ}{{{F${\cal Q}$}}}
\newcommand{\fO}{{F${\cal O}$}}
\newcommand{\afQ}{{{AF${\cal Q}$}}}
\newcommand{\afO}{{AF${\cal O}$}}

\begin{document}

\title{Landau theory of multipolar orders in Pr(TM)$_2$X$_{20}$ Kondo materials}

\author{SungBin Lee}
\email{sungbin@kaist.ac.kr}
\affiliation{Department of Physics, Korea Advanced Institute of Science and Technology, Daejeon, 34141, Korea}
\author{Simon Trebst}
\affiliation{Institute for Theoretical Physics, University of Cologne, 50937 Cologne, Germany}
\author{Yong Baek Kim}
\affiliation{Department of Physics, University of Toronto, Toronto,
Ontario M5S 1A7, Canada}
\affiliation{Canadian Institute for Advanced Research, Toronto, Ontario, M5G 1Z8, Canada}
\author{Arun Paramekanti}
\email{arunp@physics.utoronto.ca}
\affiliation{Department of Physics, University of Toronto, Toronto,
  Ontario M5S 1A7, Canada}
\affiliation{Canadian Institute for Advanced Research, Toronto, Ontario, M5G 1Z8, Canada}
  
\begin{abstract}
A series of Pr(TM)$_2$X$_{20}$ (with TM=Ti,V,Rh,Ir and X=Al,Zn) Kondo materials, containing non-Kramers Pr$^{3+}$ $4f^2$ moments on a diamond lattice, 
have been shown to exhibit intertwined orders such as quadrupolar order and superconductivity. 
Motivated by these experiments, we propose and study a Landau theory of multipolar order to capture the phase diagram and its field dependence. 
In zero magnetic field, we show that different quadrupolar states, or the coexistence of quadrupolar and octupolar orderings, may lead to ground states
with multiple broken symmetries. Upon heating, such states may undergo
two-step thermal transitions into the symmetric paramagnetic phase, with partial restoration of broken symmetries in the intervening phase.
For nonzero magnetic field, we show the evolution of these thermal phase transitions strongly depends on the field direction, due to clock 
anisotropy terms in the free energy.
Our findings shed substantial light on experimental results in the Pr(TM)$_2$Al$_{20}$ materials. We propose further experimental tests to distinguish
purely quadrupolar orders from coexisting quadrupolar-octupolar orders.
\end{abstract}

\maketitle


\section{Introduction}

Heavy fermion materials with partially filled $4f$ or $5f$ shells often exhibit unusual phases attributed to broken symmetries involving higher order multipolar degrees of freedom.
Given the challenging task of experimentally probing such broken symmetries, they are generally dubbed ``hidden orders''.\cite{Chandra:2002cz, kuramoto2009multipole, kusunose2008description,santini2009multipolar,shiina1997magnetic,kitagawa1996possible,caciuffo2003multipolar,morin1982magnetic,lee2015optical,suzuki2005quadrupolar} In order to obtain a broad understanding 
of such systems, it is useful to study families of materials which share similar 
underlying microscopics and related phenomenology. A particularly useful example is provided by the Pr(TM)$_2$X$_{20}$ intermetallic compounds, with TM=Ti,V,Rh,Ir and X=Al,Zn.\cite{sakai2012thermal,sato2012ferroquadrupolar,onimaru2016exotic,onimaru2011antiferroquadrupolar,tsujimoto2015anomalous,onimaru2012simultaneous,onimaru2010superconductivity,sakai2012superconductivity,matsubayashi2012pressure,matsubayashi2014heavy,tsujimoto2014heavy,iwasa2017evidence,u2011musr}
All these materials have been shown to exhibit quadrupolar orders and superconductivity at lower temperatures.
The common ingredient in this family is the local moment degree of freedom provided by the Pr ion. The interplay of strong spin-orbit coupling (SOC) and weaker
crystal field splitting leads to a ground state $\Gamma^{(3)}$ non-Kramers doublet on Pr, with
a significant gap to the higher order multiplets. This doublet carries no dipole moment, but has nonzero quadrupolar and octupolar moments.\cite{onimaru2016exotic} 
A key motivation to explore such 
materials was the theoretical proposal that conduction electrons scattering off such doublets would lead to non-Fermi liquid behavior 
associated with the single ion two-channel Kondo model.\cite{cox1999exotic,cox1987quadrupolar} The low temperature fate of the Kondo lattice system, however, remains an important open question. 
An understanding of these ground states is also important for clarifying the possible quantum phase transitions of these heavy fermion materials.\cite{si2010heavy,stewart1984heavy,gegenwart2008quantum,fisk1995physics}


Recent experiments on these Pr(TM)$_2$X$_{20}$ materials have confirmed the existence of quadrupolar ordering. For instance,
PrTi$_2$Al$_{20}$ displays ferroquadrupolar (\fQ) order 
below $T_{Q}\! \sim\! 2$K, while antiferroquadrupolar (\afQ) order is found in PrV$_2$Al$_{20}$ ($T_Q \!\sim\! 0.75$K), in  PrIr$_2$Zn$_{20}$ ($T_Q \!\sim\! 0.11$K), 
and PrRh$_2$Zn$_{20}$ ($T_Q \!\sim\! 0.06$K). \cite{sakai2011kondo,koseki2011ultrasonic,sato2012ferroquadrupolar,sakai2012superconductivity,onimaru2010superconductivity,onimaru2012simultaneous,iwasa2017evidence}Interestingly,
PrV$_2$Al$_{20}$ exhibits an additional phase transition at $T^* \! \sim \! 0.65$K, and shows non-Fermi liquid behavior above $T_Q$
in contrast to the Fermi liquid behavior observed in PrTi$_2$Al$_{20}$.\cite{sakai2011kondo,tsujimoto2014heavy,tsujimoto2015anomalous} This may be due to stronger hybridization between local moments
and conduction electrons in PrV$_2$Al$_{20}$, leading to proximity to an underlying quantum critical 
point.\cite{tsujimoto2014heavy,matsumoto2016strong,tokunaga2013magnetic} The precise nature of the antiferroquadrupolar orders and the 
additional transition in PrV$_2$Al$_{20}$ however remain to be understood.

Further insights into the phase diagram come from experiments studying the impact of a magnetic field.\cite{sakai2011kondo,tsujimoto2014heavy,shimura2013evidence,koseki2011ultrasonic,taniguchi2016nmr,ishii2013antiferroquadrupolar,higashinaka2017antiferroquadrupolar} For {\fQ} order, it is well known that the magnetic field couples at ${\cal O}({\boldsymbol B}^2)$ directly to the order parameter, which converts the sharp paramagnet-to-{\fQ} 
thermal transition into a crossover;
this has been observed in PrTi$_2$Al$_{20}$.\cite{sakai2011kondo} On the other hand, the multiple transitions in PrV$_2$Al$_{20}$ at $T_Q$ and $T^*$ are found to survive 
at nonzero fields, and moreover evolve in a manner which depends strongly on the field direction.\footnote{Akito Sakai, Talk given at the J-Physics Topical Meeting on 
``Exotic Phenomena in Itinerant Multipole Systems'', ISSP, University of Tokyo, December 18, 2017}

In this work, we investigate a symmetry-based Landau theory to gain insight into multipolar orders, their phase transitions, and the impact of the magnetic field, which are motivated by 
experiments on Pr(TM)2Al20 with TM=Ti,V. Given that a microscopic model of the Pr doublet, which hosts both quadrupolar and octupolar moments hybridized to conduction electrons, is likely to depend on details of the material-specific band structure and Kondo couplings, we believe such a symmetry-based approach should also be of broader relevance.

Our Landau theory includes uniform and staggered quadrupolar orders which are relevant to the {\fQ} and {\afQ} states.
For {\fQ} order, we find that the Landau theory permits a cubic anisotropy term, which was previously pointed out within a microscopic theory and 
classical Monte Carlo study of a lattice model.\cite{hattori2014antiferro,hattori2016antiferro} This
selects an {\fQ} ordered state which is consistent with experimental results on PrTi$_2$Al$_{20}$.\cite{taniguchi2016nmr}
On the other hand, {\afQ} order is generally accompanied by a ``parasitic" {\fQ} order due to a cubic term which couples them.
However, previous work found a single transition at which both orders are generated, and it thus does not explain the emergence of the two
transitions observed in PrV$_2$Al$_{20}$ at zero field. Moreover, the octupole moment carried by the doublet is typically ignored in previous studies;
however, we have argued in recent work that this might be an important ingredient, and studied an appropriate diamond lattice model 
which hosts coexisting quadrupolar and octupolar orders.\cite{freyer2018two}

Our Landau theory approach, which incorporates quadrupolar as well as octupolar order parameters, and symmetry-allowed
clock anisotropies in the free energy,
suggests two possible ways to explain the multiple thermal transitions in PrV$_2$Al$_{20}$,\cite{tsujimoto2015anomalous} and understand
the field evolution of the phase diagrams.

(i) Within a purely quadrupolar description, we show that the interplay of {\afQ} and {\fQ} orders can lead to
a second (lower temperature) transition at $T^*$ within the {\afQ} phase due to a competition between different clock
terms in the free energy. The intermediate phase in this picture preserves an Ising $S_{4z}$ symmetry, which is further broken
for $T < T^*$. 

(ii) Alternatively, we consider the more
exotic possibility that the lower temperature transition at $T^*$ might correspond
to the ordering of octupolar degrees of freedom within the {\afQ} phase, which would lead to spontaneous time-reversal symmetry breaking for
$T < T^*$.

We find that both scenarios can potentially lead to similar experimental phase diagrams and their magnetic field evolution
while the way that zero and finite temperature transitions are connected may be different in the two cases.
We therefore conclude with a discussion of possible further experimental
tests to distinguish between these two scenarios.

\section{Symmetries}

Pr(TM)$_2$X$_{20}$ (with TM=Ti,V,Rh,Ir and X=Al,Zn) are cage compounds with the space group Fd$\bar{3}$m. In particular, the Pr$^{3+}$ $4f^{2}$ ions live on 
a diamond lattice, with each ion at the center of the Frank Kasper cage formed by 16 neighboring X ions with the local point group $T_d$.\cite{onimaru2016exotic}  Strong SOC
leads to a total angular momentum $J=4$ on the Pr ion, while crystal field splitting leads to a $\Gamma_3$ doublet ground state. (We note that
PrRh$_2$Zn$_{20}$ has the local point group $T$ due to a further structural transition, and has a $\Gamma_{23}$ doublet ground state.)\cite{sakai2011kondo,onimaru2016exotic} The
$\Gamma_3$ doublet wavefunctions are given by~\cite{sato2012ferroquadrupolar,hattori2014antiferro}
\bea
\ket{\Gamma_3^{(1)}} &=& \frac{1}{2} \sqrt{\frac{7}{6}} \ket{4} - \frac{1}{2}  \sqrt{\frac{5}{3}} \ket{0} + \frac{1}{2} \sqrt{\frac{7}{6}} \ket{-4} \, \nonumber \\
\ket{\Gamma_3^{(2)}} &=& \frac{1}{\sqrt{2}} \ket{2} + \frac{1}{\sqrt{2}} \ket{-2}.
\label{eq:gamma3}
\eea
In these compounds, the first excited triplet $\Gamma_4$ or $\Gamma_5$ is  separated from the ground doublet by $\Delta \approx 30$-$70$K. This allows us to study
the broken symmetry phases, which typically have transition temperatures $\lesssim 5$K, by projecting to the $\Gamma_3$ (or $\Gamma_{23}$) doublets.
Using these doublets, we define pseudospin-1/2 basis as in Ref.~[\onlinecite{freyer2018two}], namely,
\bea
\ket{\upa} &\!\equiv \!& \frac{1}{\sqrt{2}} (\ket{\Gamma_3^{(1)}} \!+\! i \ket{\Gamma_3^{(2)}}) \\
\ket{\dna} &\!\equiv\!& \frac{1}{\sqrt{2}} (i \ket{\Gamma_3^{(1)}}\! +\!\ket{\Gamma_3^{(2)}}).
\eea
We identify the corresponding pseudospin operators in terms of Stevens operators \cite{stevens1952matrix,lea1962raising}
$O_{22} \!=\! \frac{\sqrt{3}}{2} (J^2_x \!-\! J^2_y)$, $O_{20} \!=\! \frac{1}{2}(3 J^2_z\!-\!J^2)$, and 
$T_{xyz} \!=\! \frac{\sqrt{15}}{6} \overline{J_x J_y J_z}$ (with the overline denoting a fully symmetrized product),
as 
\be
\tau^x = - \frac{1}{4} O_{22};~\tau^y = -\frac{1}{4} O_{20};~\tau^z = \frac{1}{3\sqrt{5}} T_{xyz}.
\ee
Here, the components of the pseudospin $\vec\tau$ are such that $(\tau^{x},\tau^y) \!\equiv\! \vec \tau^\perp$
describes a time-reversal invariant quadrupolar moment, while $\tau^z$ describes a time-reversal odd octupolar 
moment. 

The point group symmetries of Pr$^{3+}$ ions include $\mathcal{S}_{4z}$ ($\pi/2$ rotation about $z$ axis and inversion about a site), $\mathcal{C}_{31}$ ($2\pi/3$ rotation along (111) direction), $\mathcal{\sigma}_{d1}$(mirror reflection with a plane perpendicular to (1$\bar{1}$0) direction) and $\mathcal{I}$ (bond-centered inversion). Under these point group operations 
and time reversal ($\Theta$), the pseudospins transform as:
\bea
{\Theta} &:&~~ \tau_{A/B}^z \rightarrow -~\tau_{A/B}^z , \\
\mathcal{I} &:&~~ \vec \tau_A ~~~ \leftrightarrow ~~~ \vec \tau_B , \\
\mathcal{S}_{4z} & :&~~ \tau_{A/B}^\pm \rightarrow -~ \tau^\mp_{A/B};~ \tau_{A/B}^z \rightarrow -~ \tau^z_{A/B}\\
\mathcal{\sigma}_{d1} &:&~~ \tau_{A/B}^\pm \rightarrow -~\tau^\mp_{A/B};~ \tau_{A/B}^z \rightarrow -~ \tau^z_{A/B}\\
\mathcal{C}_{31} & :&~~ \tau_\mu^\pm ~~~\rightarrow ~~~e^{ \pm i 2\pi/3} \tau^\pm_\mu
\eea
Since the pseudospins transform in the same manner under $S_{4z}$ and $\sigma_{d1}$, we drop the
$\sigma_{d1}$ symmetry in the following analysis. We next use these symmetries in order to construct the Landau theory.

\section{Landau theory}

In this paper, we study the simplest scenarios with uniform or two-sublattice orders which do not enlarge the unit cell of the diamond lattice. Thus, we consider
Ferro$\mathcal{Q}$uadrupole ({\fQ}), AntiFerro$\mathcal{Q}$uadrupole ({\afQ}), Ferro$\mathcal{O}$ctupole ({\fO}) and AntiFerro$\mathcal{O}$ctupole ({\afO}) broken
symmetry states. Some of these orders could potentially coexist. Let us introduce uniform and staggered multipolar
order parameters
\bea
 { \phi}_{u,s} &\!\equiv\!& \la \tau_A^+ \ra \pm \la \tau_B^+ \ra \\
 m_{u,s} &\!\equiv\!&  \la \tau^z_A \ra \pm \la \tau^z_{B} \ra \,.
 \label{eq:order-parameters}
 \eea
Here, the complex scalars $\phi_{u,s}$ denote, respectively, the {\it u}niform (for \fQ) and {\it s}taggered parts (for \afQ) of the $XY$ quadrupolar order, while
the real scalars $m_{u,s}$ refer to the {\it u}niform (for \fO) and {\it s}taggered parts (for \afO) of the Ising octupolar order.
The underlying crystal and time-reversal symmetry transformations act on the order parameters $\phi_{u,s}$ and $m_{u,s}$ as follows:
\bea
\!\! \!\! \!\! {\Theta} &:& \phi_{u,s} \to \phi_{u,s};~m_{u,s} \to -m_{u,s} \\
\!\! \!\! \!\! \mathcal{I} &:& (\phi_u,m_u) \to (\phi_u,m_u);~(\phi_s,m_s) \to - (\phi_s,m_s)\\
\!\! \!\! \!\! \mathcal{S}_{4z} & : & \phi_{u,s} \to -\phi^*_{u,s};~m_{u,s} \to -m_{u,s}\\
\!\! \!\! \!\! \mathcal{C}_{31} & : & \phi_{u,s} \to e^{i 2\pi/3} \phi_{u,s};~m_{u,s} \to m_{u,s} \,.
\label{eq:symmetry-orderparameters}
\eea
The symmetry-allowed terms in the Landau free energy with independent order parameters are thus:
\bea
\!\! \!\! {\cal F}_{\phi u} &\!=\!& r_{u\phi} |\phi_u|^2 \!+\! i v (\phi_u^3 \!-\! \phi_u^{*3})  \!+\! g_{u\phi} |\phi_u|^4  \!+\! \ldots
\label{eq:F_phiu}  \\
\!\! \!\! {\cal F}_{\phi s} &\!=\!& r_{s\phi} |\phi_s|^2  \!+\! g_{s\phi} |\phi_s|^4  \!+\! w (\phi_s^6  \!+\! \phi_s^{*6})  \!+\! \ldots
\label{eq:F_phis} \\
\!\! \!\! {\cal F}_{m u} &\!=\!& r_{u m} m_u^2  \!+\! g_{u m } m_u^4 \!+\! \ldots
\label{eq:F_mu} \\
\!\! \!\! {\cal F}_{m s} &\!=\!& r_{s m} m_s^2  \!+\! g_{s m} m_s^4 \!+\! \ldots \,,
\label{eq:F_ms}
\eea
where the ellipses denote dropped higher order terms.
The important difference between the {\fQ} versus {\afQ} free energies appears in the ``clock'' anisotropy terms which break $XY$ symmetry
for $\phi_u, \phi_s$ respectively; this is cubic for {\fQ}
and sixth order for {\afQ}. This free energy must be supplemented by ${\cal F}_{\rm int}$ which encapsulates interactions between the 
different order parameters. Symmetry allows for a single cubic interaction, 
\be
{\cal F}_{\rm int}^{(3)}= i \lambda (\phi_s^2 \phi_u - 
\phi_s^{*2} \phi^*_u).
\ee 
This leads to ``parasitic'' {\fQ} order $\phi_u \! \sim \! \phi_s^{*2}$ in an {\afQ} state.
Additional quartic interactions between order parameters take the form
\bea
\!\!\!\!\!\!  {\cal F}_{\rm int}^{(4)} &=&  c_1 |\phi_u|^2 |\phi_s|^2 + c_2 m_u^2 m_s^2 + c_3 |\phi_u|^2 m_u^2 \nonumber \\
&+& c_4 |\phi_s|^2 m_s^2  + c_5 |\phi_u|^2 m_s^2 + c_6 |\phi_s|^2 m_u^2 \,.
\label{eq:F_int}
\eea
Such terms can lead to coexistence of quadrupolar and octupolar order parameters depending on the signs of the coefficients.
Below, we will analyze this Landau free energy in various cases, starting from the simplest example.

\subsection{{\fQ} order in PrTi$_2$Al$_{20}$}

PrTi$_2$Al$_{20}$ exhibits {\fQ} order, so we can focus on the single term ${\cal F}_{\phi u}$ in Eq.~\eqref{eq:F_phiu} above.\cite{sakai2011kondo,koseki2011ultrasonic,sato2012ferroquadrupolar} For $r_{u\phi} > 0$,
this describes a paramagnetic (PM) phase with $\phi_u=0$, while $r_{u\phi} < 0$ leads to {\fQ} order with $\phi_u \neq 0$. 
The phase of $\phi_u \equiv |\phi_u| {\rm e}^{i\theta_u}$ 
is determined by the clock term $v$. For $v > 0$, we favor $\theta_u = \pi/6 + 2 n \pi/3$ (with integer $n$), while $v < 0$ pins $\theta_u = \pi/6 + (2 n+1) \pi/3$.
In particular, either sign of $v$ favors $O_{20}$ order over $O_{22}$ order, which is consistent with nuclear magnetic resonance (NMR) experiments \cite{taniguchi2016nmr}
on PrTi$_2$Al$_{20}$. In the ``hard-spin'' limit, the theory for the PM-to-{\fQ} transition is a $Z_3$ clock model which is known to exhibit
 a first-order transition
in three dimensions (3D).\cite{wu1982potts, hove2003criticality} 
However,  disorder effects have been shown in certain examples to convert first-order transitions into continuous phase transitions.\cite{Bellafard2015}
Such effects may be important in understanding experimental observations; this needs further investigation.

\subsection{{\afQ} with `parasitic' {\fQ} order}

Let us ignore the octupolar orders $m_{u},m_{s}$, and focus on the free energy
${\cal F}_{\phi u} + {\cal F}_{\phi s} + {\cal F}_{\rm int}^{(3)} + c_1 |\phi_u|^2 |\phi_s|^2$. For an {\afQ} transition driven by $r_{s\phi} < 0$,
we get $\phi_s \neq 0$. This {\afQ} transition will happen within mean field theory at $T_Q$ if we set $r_{s\phi} = \alpha_s (T - T_Q)$, with
$\alpha_s > 0$. 
In this case, even if $r_{u\phi} > 0$, the cubic interaction $\lambda \neq 0$ in ${\cal F}_{\rm int}^{(3)}$ leads
to $\phi_u \neq 0$. 
It is useful to begin our analysis of the interplay of {\afQ} and {\fQ} orders 
by considering the regime where $r_{u\phi}$ is large. The resulting {\fQ} order is then parasitic, and 
it will be slaved to the {\afQ} order.
Let us simplify the problem by setting $(v, g_{u\phi}) \to 0$ to leading order, and minimizing the free energy with respect to $\phi_u$ which leads to
\be
\phi_u = \frac{i \lambda}{r_{u\phi}} \phi_s^{*2} \,.
\ee
Substituting back, the full free energy is given by
\bea
\!\!\! \!\!\! {\cal F}^{\rm eff}_{\phi s}  &=& r_{s\phi} |\phi_s|^2  + g^{\rm eff}_{s\phi}  |\phi_s|^4 + w^{\rm eff} (\phi_s^6 + \phi_s^{*6})  + \ldots \,,
\eea
where 
\bea
g^{\rm eff}_{s\phi}  &=& g^\pdg_{s\phi} - \frac{\lambda^2}{r_{u\phi}} \\
w^{\rm eff}  &=& w + v \frac{\lambda^3}{r_{u\phi}^3} \,.
\eea
With $\phi_s = |\phi_s| {\rm e}^{i\theta_s}$,
we find that the clock term with $w^{\rm eff} > 0$ favors $\theta_s = (2 n+1) \pi/6$, while $w^{\rm eff} < 0$ would favor $\theta_s = 2 n \pi/6$.
Now, even if $r_{u\phi} > 0$, it may have a temperature dependence as $r_{u\phi} = r_{u\phi} (0) + \alpha_u T$ with $r_{u\phi} (0) > 0, \alpha_u > 0$. 
Such a (benign) temperature dependence of $r_{u\phi}$ could, nevertheless, lead to a
change of sign of $g^{\rm eff}_{s\phi}$ which could lead to first-order transitions, or a sign change of $w^{\rm eff}$ (if the product $w v \lambda < 0$) which may modify the
competition between the different clock terms.
This, admittedly crude, argument suggests that the interplay of {\afQ} and {\fQ} orders could lead to a rich phase diagram with new phases and phase
transitions.

\begin{figure}[t]
\subfloat[]
{\includegraphics[width=0.22 \textwidth]{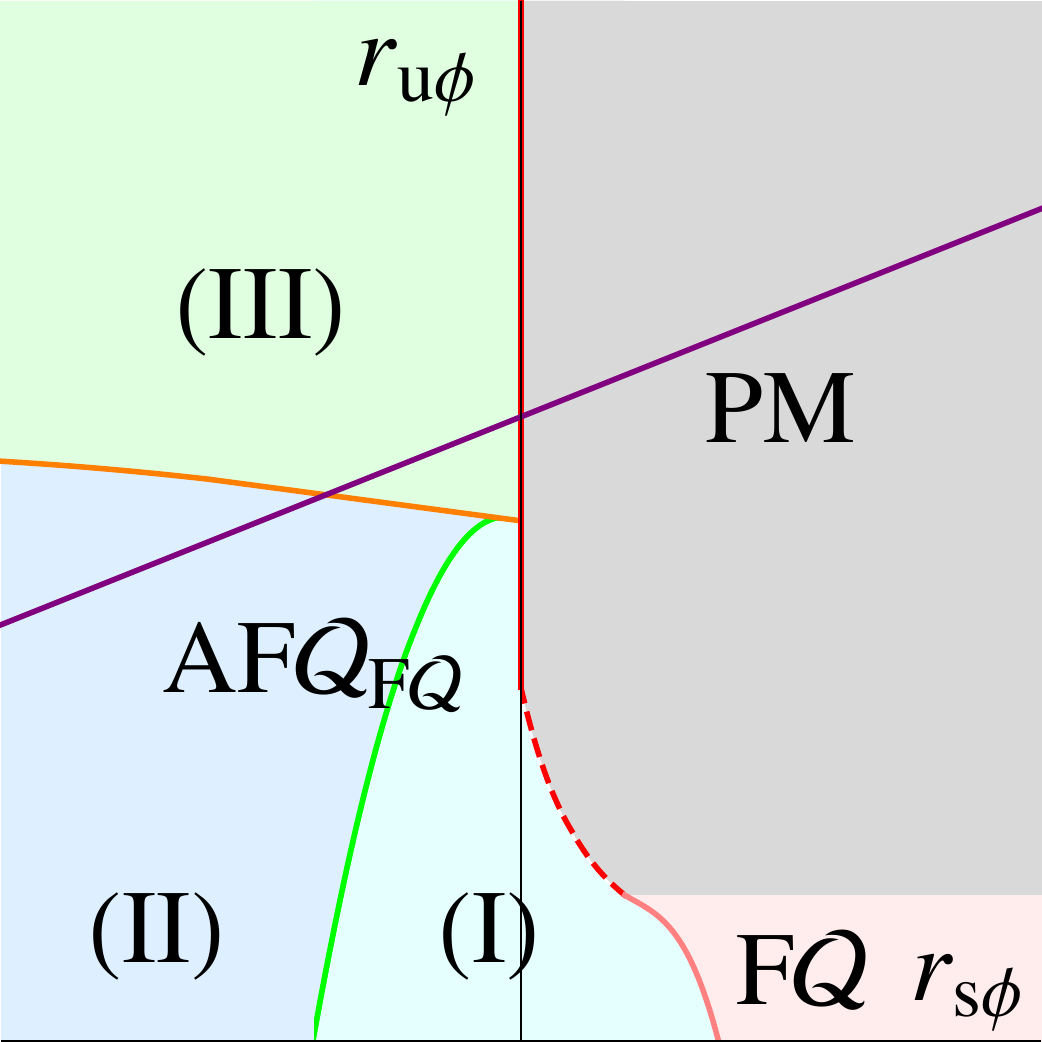}}
\quad
\subfloat[]
{\includegraphics[width=0.22 \textwidth]{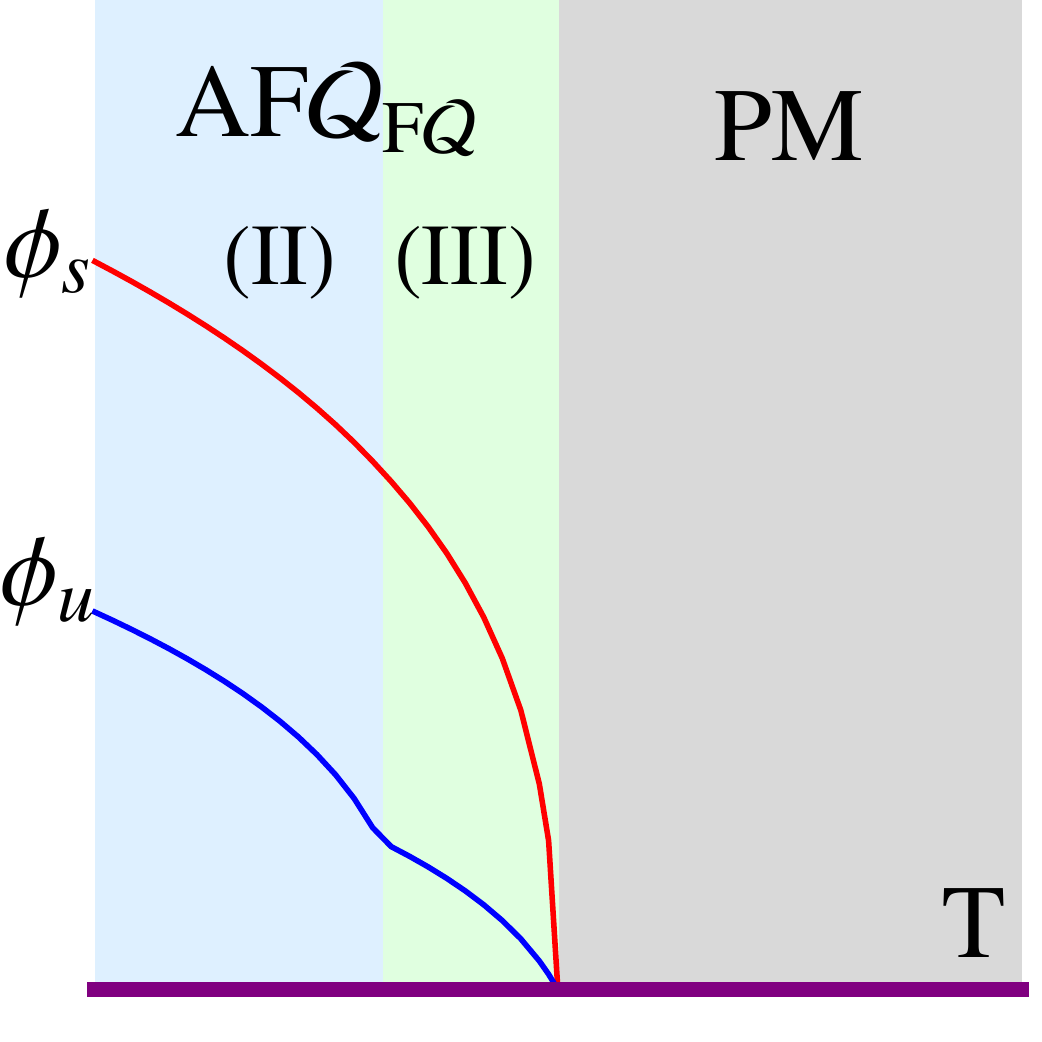}}
\vspace{0.1pt}
\subfloat[]
{\includegraphics[width=0.5 \textwidth]{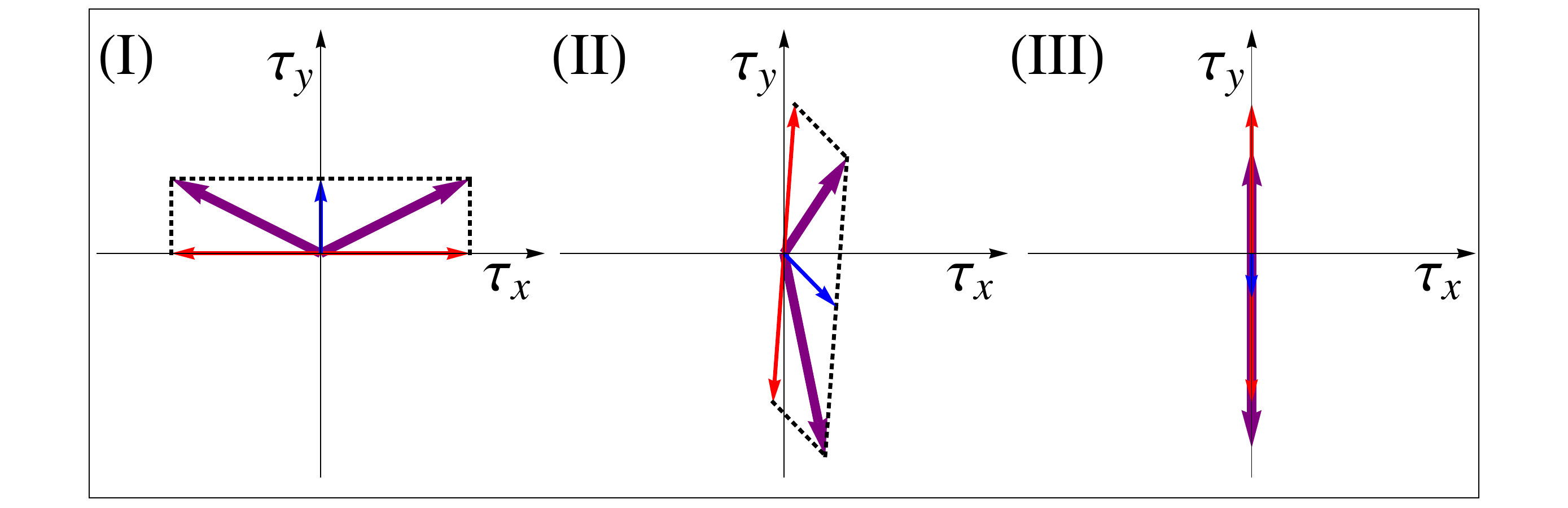}}
\caption{[Color online] (a) Phase diagram of the Landau theory described by
${\cal F}_{\phi u}+{\cal F}_{\phi s}+ {\cal F}^{(3)}_{\rm int}$ with {\afQ} and {\fQ} order parameters as functions of $r_{s \phi}$ and $r_{u \phi}$. Here, we take $w v \lambda \!<0$. The various phases are paramagnet (PM), {\fQ} and three distinct phases (I),(II) and (III) for {\afQ} with parasitic {\fQ} ({\afQ}$_{\text{\fQ}}$). See main text for details. (b) Plot of the order parameters as a function of temperature T, a cut through the trajectory (purple line) in panel (a). Red and blue lines represent the magnitude of order parameters $\phi_s$ and $\phi_u$. (c) Common origin plots of distinct {\afQ}$_{\text{\fQ}}$ phases (I),(II) and (III). Red and Blue arrows exhibit {\afQ} and {\fQ} respectively and purple arrow is the direction of quadrupolar order resulting from combination of both {\afQ} and  {\fQ}. All these spin configurations have three-fold degeneracies with $2\pi/3$ rotation in $\tau_x$-$\tau_y$ plane.  }
\label{fig:QuadPD}
\end{figure}

In order to examine this scenario, we numerically minimize the Landau free energy ${\cal F}_{\phi s} + {\cal F}_{\phi u} + {\cal F}^{(3)}_{\rm int}$,
as a function of $r_{u\phi}$ and $r_{s\phi}$, while keeping $r_{u\phi} > 0$. For illustrative purposes, we fix $g_{u\phi}=1$ and $g_{s\phi}=1/2$,
and consider the choice for the coefficients of the clock terms $(w,v,\lambda) \equiv (1/4,-1/4,1/4)$. The resulting
phase diagram is shown in Fig.~\ref{fig:QuadPD}(a), and exhibits five different phases:
a paramagnet (PM), a {\fQ} state driven by 
the cubic term $v$, and three types of {\afQ} phases (with coexisting {\fQ} order) which result from
competition between the different clock terms in the free energy.
Fig.~\ref{fig:QuadPD}(c) shows the nature of the different {\afQ} phases, which are distinguished by the behavior of the quadrupole moment
on the two sublattices. 

In {\afQ}-I, the staggered quadrupolar order points along $\tau_x$ (O$_{22}$) while the parasitic uniform component
points along $\tau_y$ (O$_{20}$). This phase minimizes the clock anisotropy terms $v$ and $\lambda$. The {\afQ}-I state
depicted in Fig.~\ref{fig:QuadPD}(c) preserves $S_{4z}$ and $\Theta$ symmetries.

In {\afQ}-III, both the staggered and 
uniform components favor O$_{20}$ order, so the overall magnitude of the
ordered quadrupole moment is different on the two sublattices. This phase minimizes the clock terms $w$ and $\lambda$. Again,
the {\afQ}-III state
depicted in Fig.~\ref{fig:QuadPD}(c) preserves $S_{4z}$ and $\Theta$ symmetries.

Finally, {\afQ}-II is a ``frustrated'' phase, where the competition of the different clock terms $(w,v,\lambda)$ results in none of them 
being fully  minimized. This phase exhibits a generically complex superposition of O$_{20}$ and O$_{22}$ orders, with unequal magnitude
of the ordered moment on the two sublattices, and only preserves $\Theta$. We thus expect the {\afQ}-II state, which breaks the residual $S_{4z}$ 
symmetry, and thus 
has lower symmetry than {\afQ}-I or
{\afQ}-III, to arise from either one of them upon cooling.

We find that different choices for these clock coefficients, keeping the product $w v \lambda < 0$ yield phase diagrams with the same
phases and a roughly similar topology.
For instance, when we decrease $\lambda = 1/16$, we find the following differences:
(i) the {\afQ}-I phase shrinks, (ii) the {\afQ}-II to {\afQ}-III phase transition becomes first order, and (iii) there is no direct transition from 
PM into {\afQ}-I.

To see how this $(r_{s\phi},r_{u\phi})$ phase diagram might translate into a phase diagram as a function of temperature,
consider a cut through Fig.~\ref{fig:QuadPD}(a) at large $r_{u\phi}$. Such a cut will yield a PM to {\afQ}-III transition, i.e., a {\it single} transition
into a phase with coexisting {\afQ} order and parasitic {\fQ} order. This scenario is
consistent with what has been previously explored by Hattori and Tsunetsugu.\cite{hattori2014antiferro,hattori2016antiferro} 

However, for smaller $r_{u\phi}$, along the cut shown in Fig.~\ref{fig:QuadPD}(a), we find that the transition splits into two
transitions, a PM to {\afQ}-III transition, and a subsequent  {\afQ}-III to {\afQ}-II transition. 
Fig.~\ref{fig:QuadPD}(b) shows the evolution of the order parameters with ``temperature'',
where going along the cut from PM to {\afQ}-III to {\afQ}-II is viewed as corresponding to decreasing temperature.
The two thermal transitions in this scenario might potentially
explain the two observed zero field thermal transitions in PrV$_2$Al$_{20}$.\cite{tsujimoto2015anomalous,tsujimoto2014heavy} We note that while there are many possible cuts we could take which
would lead to multiple thermal transitions, the one we have chosen seems most promising from the point of view of understanding the
magnetic field evolution as discussed in Section IV. 


\subsection{Coexisting {\afQ} and octupolar orders}
Finally, let us turn to the most interesting possibility, that the two thermal transitions in PrV$_2$Al$_{20}$ correspond, respectively, to the onset of {\afQ}
and of octupolar order which spontaneously breaks time-reversal symmetry. In previous work, we have considered this possibility within a 
particular (phenomenological)
microscopic Hamiltonian with competing two-spin and four-spin interactions which we studied using classical Monte Carlo simulations.\cite{freyer2018two} Here, we
revisit this scenario using Landau theory which goes beyond a specific microscopic model. We note the precise type of octupolar order, 
either ferrooctupolar or antiferrooctupolar, does not change our Landau theory analysis performed below; without loss of generality, we thus consider 
the case with ferro-octupolar order. This distinction will of course be important when we turn in the end to a discussion of experimental consequences.


\begin{figure}[t]
\subfloat[]
{\includegraphics[width=0.22 \textwidth]{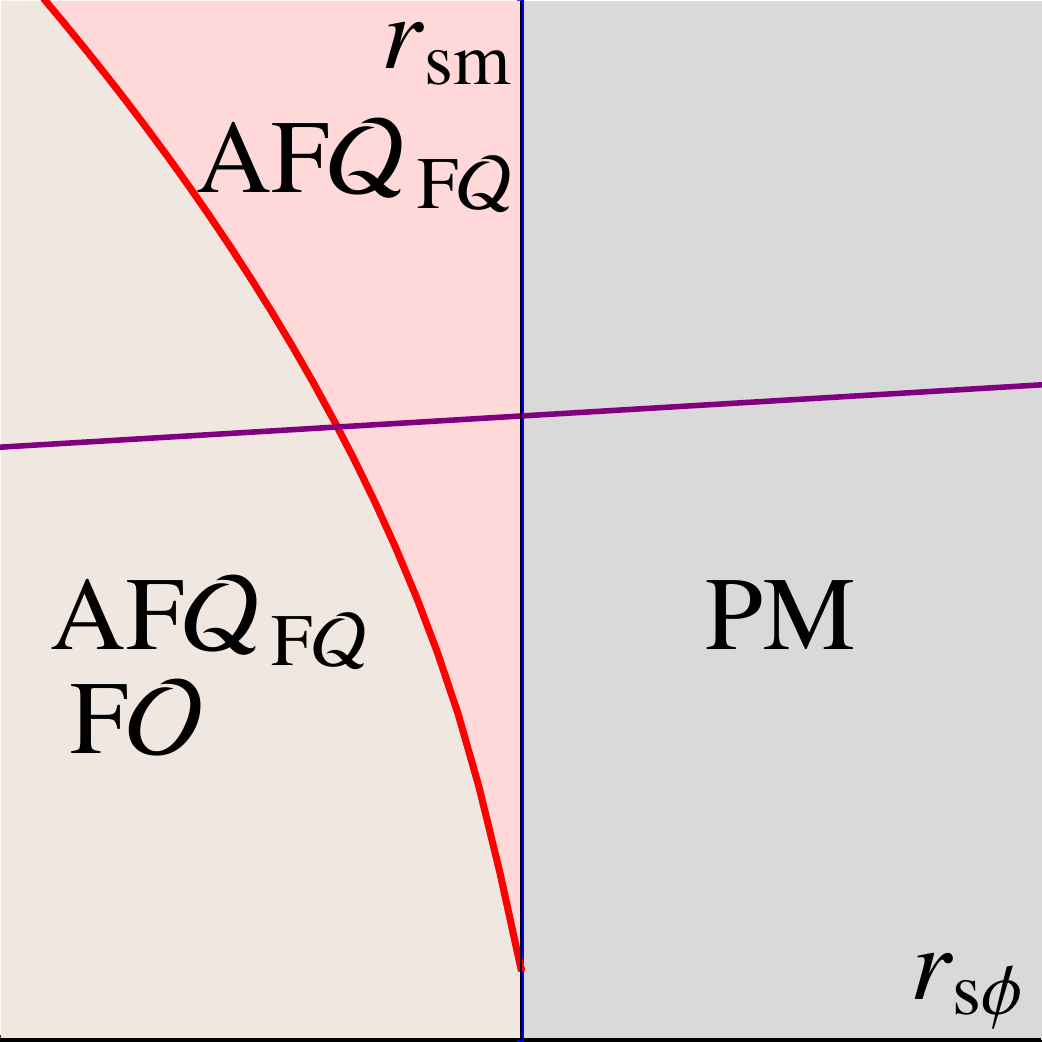}}
\quad
\subfloat[]
{\includegraphics[width=0.22 \textwidth]{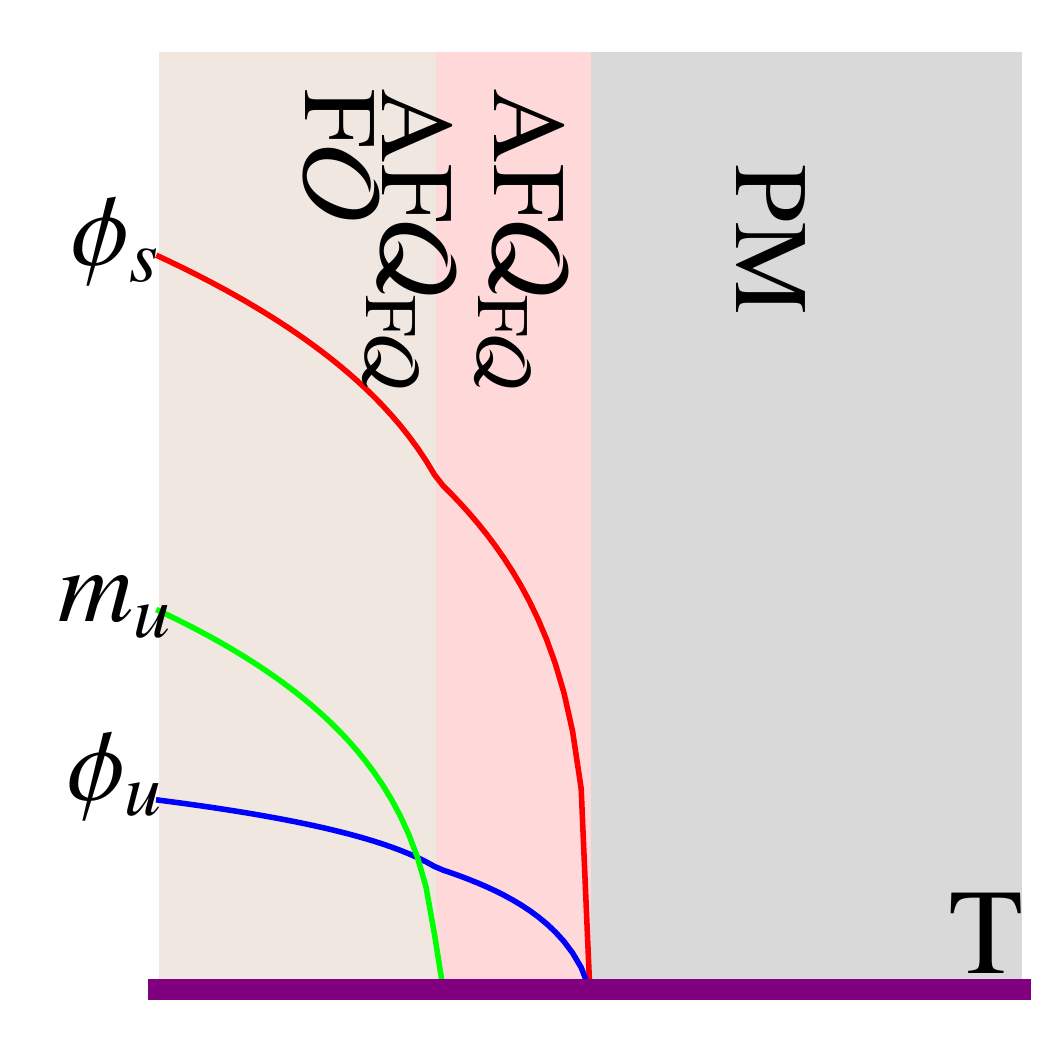}}
\vspace{1pt}
\subfloat[]
{\includegraphics[width=0.45 \textwidth]{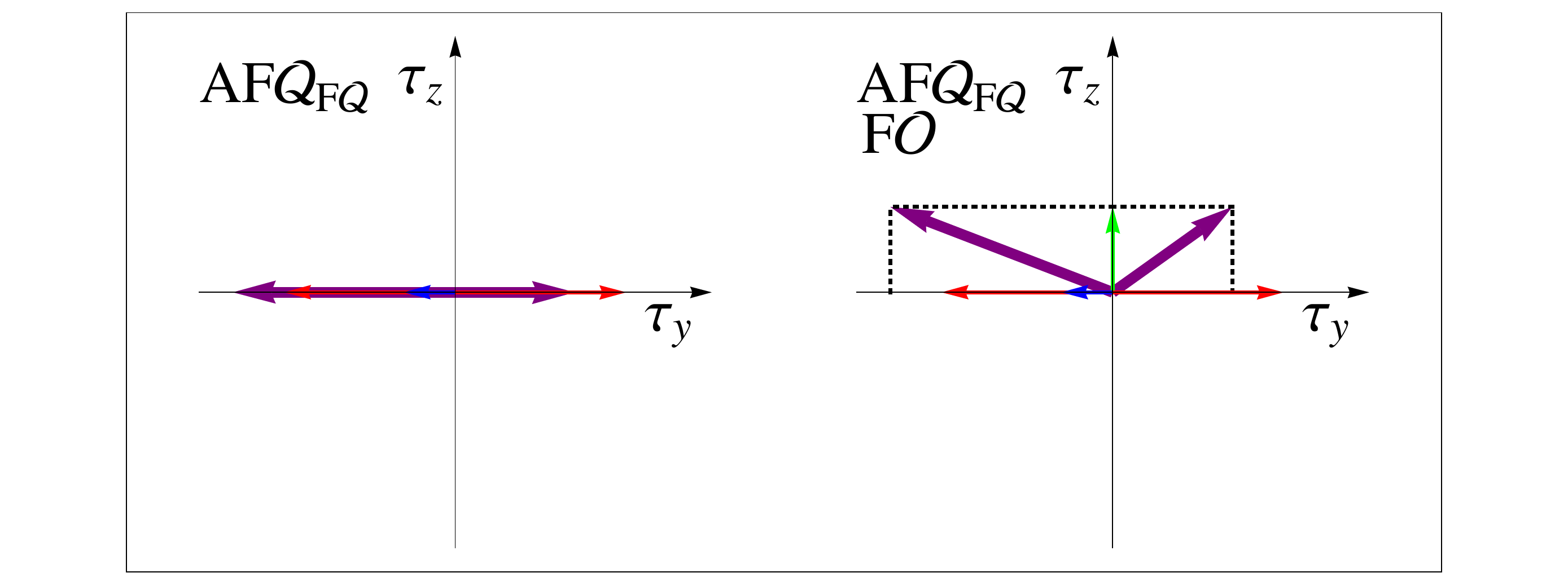}}
\caption{[Color online] (a) Phase diagram of the Landau theory described by
${\cal F}_{\phi u}\!+\!{\cal F}_{\phi s}\!+\!{\cal F}_{m u}\!+\! {\cal F}^{(3)}_{\rm int}\!+\! {\cal F}^{(4)}_{\rm int}$ with {\afQ}, {\fQ} and {\fO} order parameters as functions of $r_{s \phi}$ and $r_{s m}$. Here, we set $w v \lambda \!>0$ distinct with the case depicted in Fig.~\ref{fig:QuadPD}, thus the phase transition only arise from developing additional octupolar order. In this case, three phases exist; paramagnet (PM), {\afQ} with parasitic {\fQ} ({\afQ}$_{\text{\fQ}}$) and coexisting {\afQ} and {\fO} with parasitic {\fQ} order ({\afQ}$_{\text{\fQ}}$ {\fO}). 
See main text for details. (b) Plot of the order parameters along shown trajectory (purple line) in panel (a). Red, blue and green lines represent the magnitude of order parameters 
$\phi_s$ and $\phi_u$ and $m_u$. (c) Common origin plots of each phase.  Red, blue and green arrows exhibit magnitudes of {\afQ}, {\fQ} and {\fO} phases respectively and purple arrow is the combination of them, determining the direction of pseudospin ${\boldsymbol \tau}$. All these spin configurations have three-fold degeneracies with $2\pi/3$ rotation in $\tau_x$-$\tau_y$ plane. (Here we chose the quadrupole order configuration having only 
$\tau_y$ component.) }
\label{fig:QuadOctPD}
\end{figure}



To illustrate this interplay of {\afQ} and octupolar orders, Fig.~\ref{fig:QuadOctPD}(a) shows a phase diagram 
obtained using the Landau free energy ${\cal F}^{\rm eff}_{\phi s}+ {\cal F}_{m u}+{\cal F}^{(4)}_{\rm int}$, where we
consider having integrated out $\phi_u$ and assumed large $r_{u\phi}$ so any multiple thermal transitions must arise
from additional octupolar order. We pick
$c_6 \neq 0$ in ${\cal F}^{(4)}_{\rm int}$ in Eq.~\eqref{eq:F_int}; specifically, we chose $c_6 < 0$ to allow for a coexistence phase. 
As we vary $r_{s\phi},r_{um}$, there exist four distinct phases: a paramagnet (PM) ($\phi_s\!=\!\phi_u\!=\!m_s \!=\!0$), an {\afQ} phase with parasitic {\fQ} order
($\phi_s \! \neq \!0, \phi_u \! \neq \! 0, m_u \! = \! 0$), 
an {\fO} phase ($\phi_s \!= \! \phi_u \! =\! 0, m_u \neq 0$), and finally a phase with coexisting {\afQ} and {\fO} orders with parasitic {\fQ} order 
($\phi_s \! \neq \!0, \phi_u \! \neq \! 0, m_u \! \neq \! 0$).
Fig.~\ref{fig:QuadOctPD}(b) shows the temperature dependence of the order parameters as we ``cool'' from the PM into the phase with coexisting {\afQ} and {\fO} orders;
for simplicity, we consider going along the trajectory indicated in Fig.~\ref{fig:QuadOctPD}(a), i.e., keeping $r_{sm}$ fixed and varying $r_{s\phi}$. This clearly shows
the double transition, with the upper transition $T_Q$ being associated with {\afQ} order (with parasitic {\fQ}) and the lower transition at $T^*$ arising from the octupolar order. Fig.~\ref{fig:QuadOctPD}(c) shows the common origin plots of pseudospin ${\boldsymbol \tau}$ for {\afQ} and {\afQ}-{\fO} respectively (both with parasitic {\fQ}). 
 
\section{Impact of a magnetic field}
 
We next consider the impact of an applied magnetic field ${\boldsymbol B}$ on the Landau free energy, and its phases and phase transitions. 
The leading term is a quadratic-in-field coupling to the quadrupolar order; 
microscopically, this arises via second order perturbation theory in ${\boldsymbol B}~\cdot~{\boldsymbol J}$, where ${\boldsymbol J}$
is the $J=4$ angular momentum operator. Projecting to the $\Gamma_3$ doublet, we arrive at the form\cite{hattori2014antiferro}
\bea
H_{\text{field}} \!&=&\! \gamma B^2 (b_1 \tau^x \!+ \! b_2 \tau^y ) 
\label{eq:H_field}
\eea 
where $b_1 \! \equiv \! \frac{\sqrt{3}}{2} (\hat{b}_x^2 - \hat{b}_y^2 )$, $b_2 \! \equiv \!  \frac{1}{2} ( 3 \hat{b}_z^2 -1)$, and
$(\hat{b}_x,\hat{b}_y,\hat{b}_z)$ describes the unit vector pointing along ${\boldsymbol B}$.
The coupling constant 
\be
\gamma \!\propto\! \Big( \! -\frac{14}{3 \Delta (\Gamma_4)} \!+\! \frac{2}{\Delta (\Gamma_5) } \Big),
\ee 
with $\Delta (.)$ being
the energy of the indicated higher energy crystal field multiplets.\cite{freyer2018two}

Note that a magnetic field along the (111) direction does not directly couple to the quadrupolar moment, but even along this direction
$B^2$ could 
couple to the energy density via $|\phi_s|^2$ or $|\phi_u|^2$, with the coupling to $|\phi_u|^2$ being less important if the {\fQ} order is parasitic and small.
Moreover, along this special (111) direction, the magnetic field can couple to the octupolar moment at cubic order in the field as 
$\sim |B|^3 \hat{b}_x \hat{b}_y \hat{b}_z \tau_z$; 
however, given that this last term is expected to be much weaker for typical fields, we omit it in
the analysis below.

To proceed, it is useful to define a complex scalar $\psi_B \!\equiv \! b_1  \!+ \! i b_2$ representing the external magnetic field, which
transforms identical to the {\fQ} order parameter $\phi_u$, and thus couples to it linearly. This leads to 
terms in the Landau free energy
\bea
\!\!\!\!\!\!\!\! {\cal F}_B \!&=&\!  \gamma B^2 (\psi^*_B \phi^\pdg_u \!+\! \phi^*_u \psi^\pdg_B )
 \!+\! B^2 (\tilde{r}_{s B}  |{\phi_s} |^2 \!+\! \tilde{r}_{u B} | {\phi_u} |^2),\,
\label{eq:F_Bphis}
\eea
where we have included extra, symmetry allowed, couplings $\tilde{r}_{s B}, \tilde{r}_{u B}$ to the energy density as discussed above. Along
key high symmetry directions, $\psi_B(111) = 0$, $\psi_B(100)= {\rm e}^{i\pi/2 + i 2 n \pi/3}$, $\psi_B(110)= \frac{1}{2} {\rm e}^{-i\pi/2+ i 2n\pi/3}$.

\subsection{{\fQ} order in PrTi$_2$Al$_{20}$}

As seen from the coupling in ${\cal F}_B$ above, the direction of the magnetic field pins the quadrupolar moment direction, thus explicitly breaking
the $Z_3$ symmetry associated with the choice of phase of $\phi_u$. This converts the PM-{\fQ} transition into a smooth crossover for both (001)
and (110) field directions, as has also been predicted based on microscopic model studies and confirmed by
specific heat measurements on PrTi$_2$Al$_{20}$.

\subsection{{\afQ} with parasitic {\fQ} order}
\begin{figure}[t]
\subfloat[]
{\includegraphics[width=0.45\columnwidth]{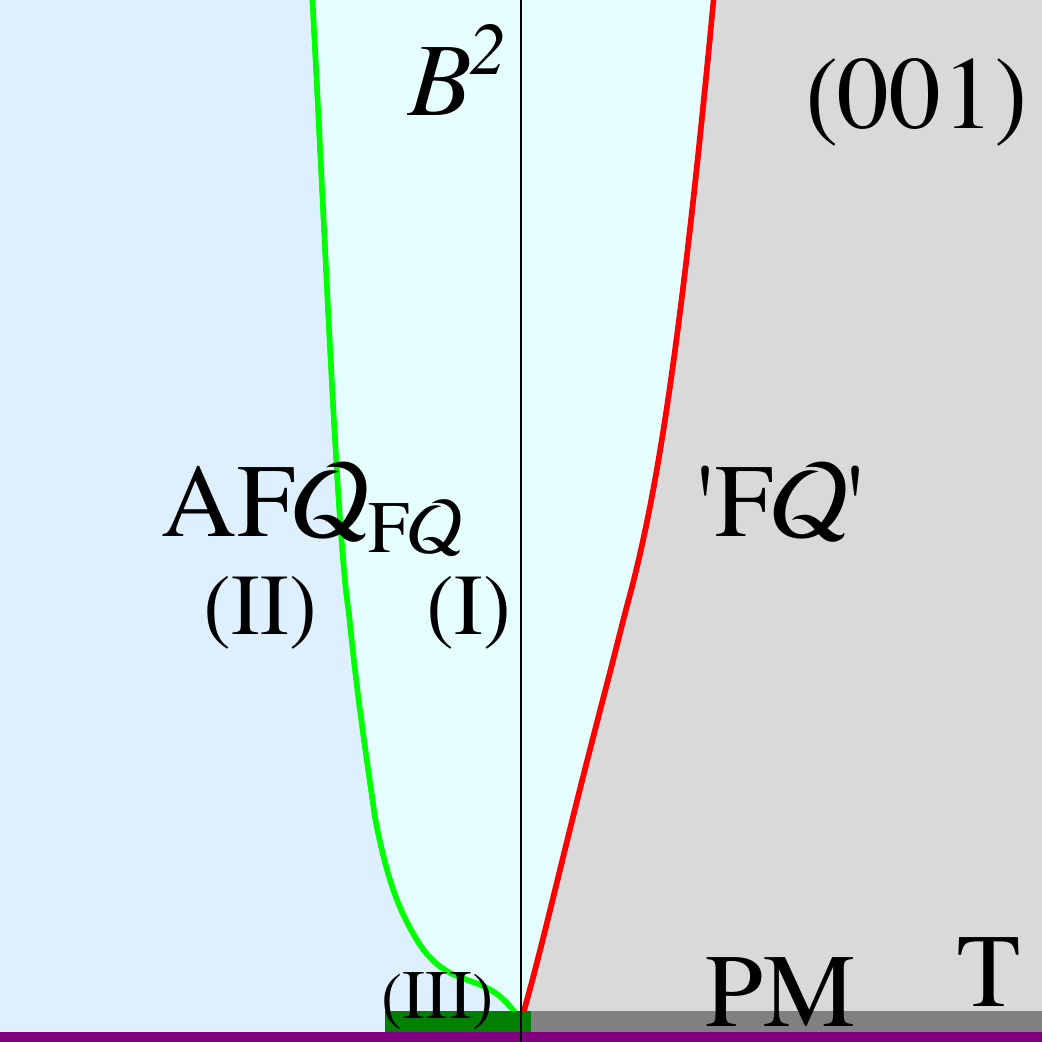}}
\qquad
\subfloat[]
{\includegraphics[width=0.45\columnwidth]{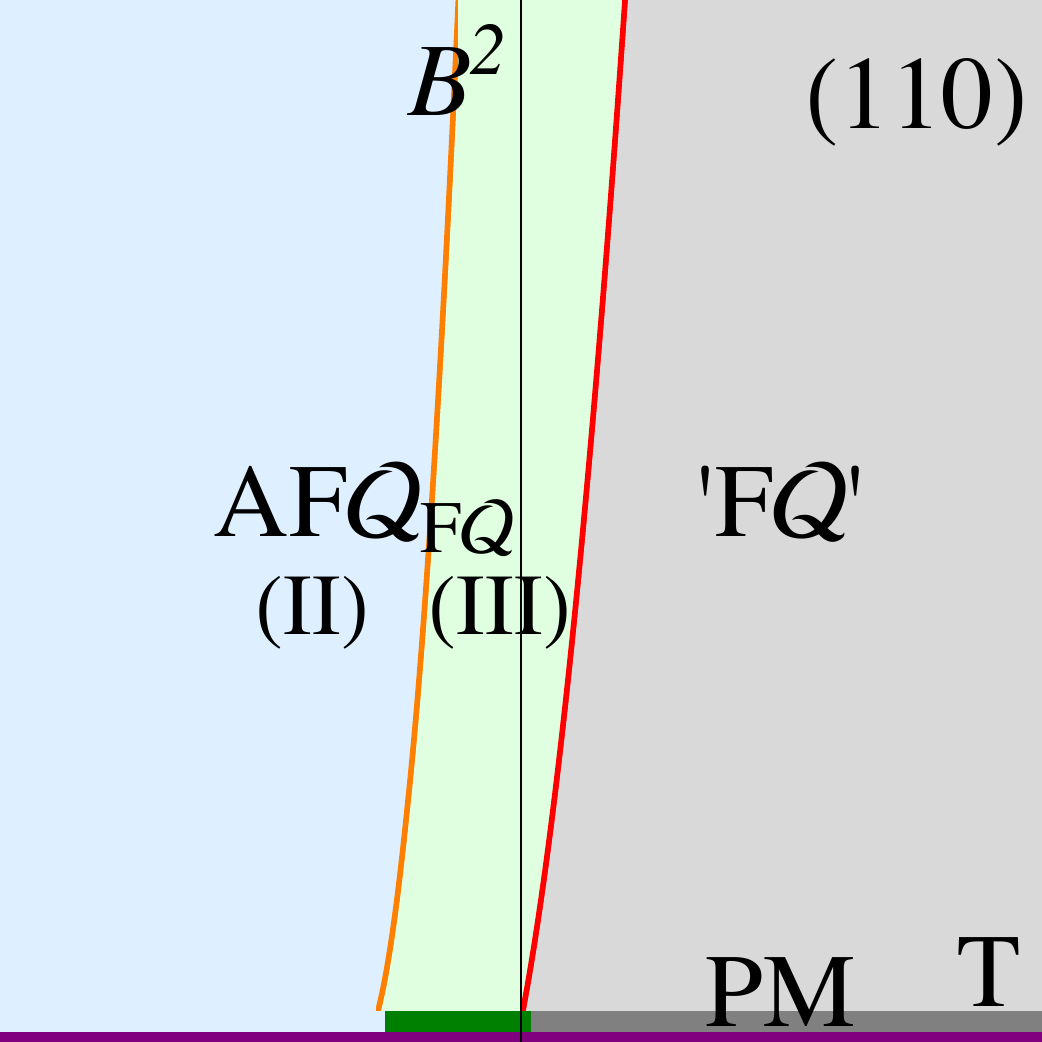}}
\caption{(a) Phase diagram of quadrupolar order as functions of magnetic field ${\boldsymbol B}//(001)$ and temperature T taking the shown trajectory along the purple line in Fig.~\ref{fig:QuadPD} (a). In the presence of field ${\boldsymbol B}//(001)$, the type (I) phase of {\afQ}$_{\text{\fQ}}$ is stabilized at intermediate temperature, whereas the type (III) phase is no longer stable with fields along (001) direction. See the main text for details.
(b) Phase diagram of quadrupolar order with ${\boldsymbol B}//(110)$. With fields, the type (III) phase is stable favored by both cubic anisotropy and field coupling of {\fQ}.}
\label{fig:PD-Quad-B}
\end{figure}

For this case, we proceed by considering the Landau free energy ${\cal F}_{\phi s} + {\cal F}_{\phi u} + {\cal F}^{(3)}_{\rm int}$ supplemented by
the field term ${\cal F}_B$. For simplicity, we set $\tilde{r}_{s B}=0$ and $\tilde{r}_{u B}=0$, and only consider the impact of the coupling $\gamma$.
Minimizing this full free energy along the cut shown in Fig.~\ref{fig:QuadPD}(a), we find the strongly direction-dependent field evolution displayed in 
Fig.~\ref{fig:PD-Quad-B} for fields along (001) and (110) directions. 
In both cases, the field couples linearly to $\phi_u$, and thus pins its phase as soon as ${\boldsymbol B}\! \neq \!0$. 
We refer to the resulting phase as `{\fQ}' to denote that it is not a symmetry broken {\fQ} state, but rather a field induced {\fQ} state which is thus
qualitatively similar to a PM.
Along the (001) direction, the entire region of {\afQ}-III and {\afQ}-II gets replaced
by the {\afQ}-II phase as $\phi_u$ cants away from pure O$_{20}$ order, while phase {\afQ}-I emerges only for nonzero $B$ from the PM-to-{\afQ}-III
transition point. Along the (110) direction however, all three phases present at zero field and the corresponding two thermal phase transitions survive even for $B \neq 0$.

\subsection{Coexisting {\afQ} and octupolar orders}
\begin{figure}[t!]
                 \subfloat[]
		{\includegraphics[width=0.45\columnwidth]{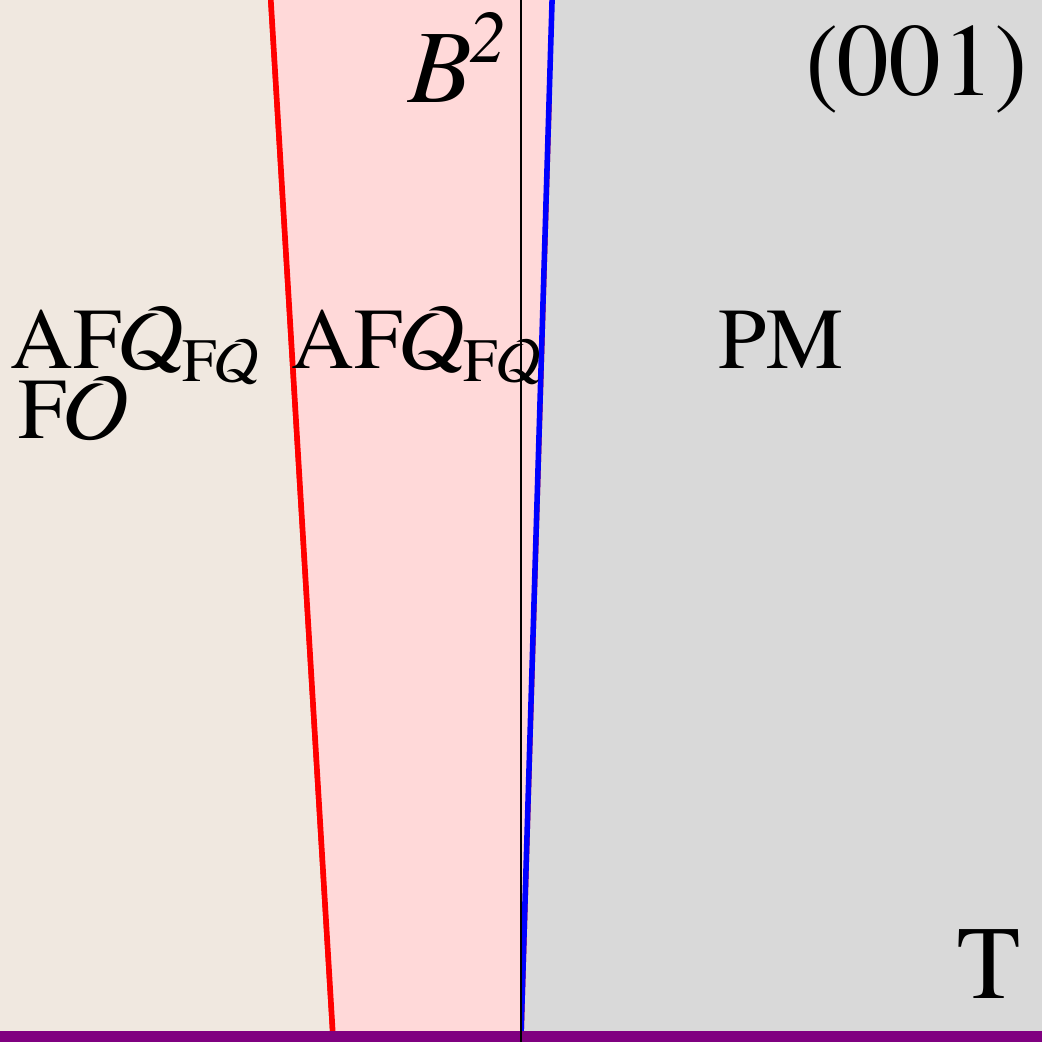}}
		\qquad
		\subfloat[]
		{\includegraphics[width=0.45\columnwidth]{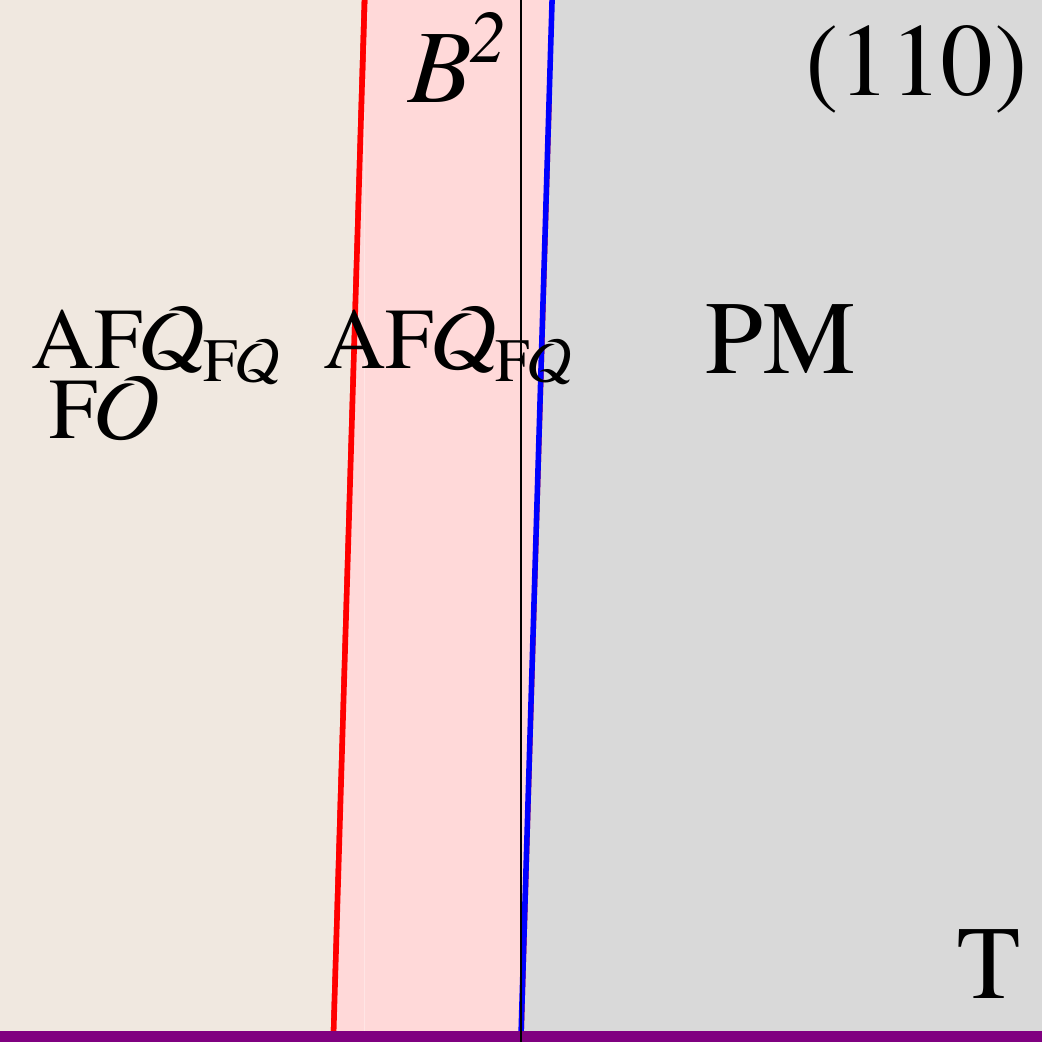}}
	\caption{(a) Phase diagram of quadrupole-octupole order as functions of magnetic field ${\boldsymbol B}//(001)$ and temperature T taking the shown trajectory along the purple line in Fig.~\ref{fig:QuadOctPD}(a). With fields along (001) direction, the phase transition temperature between {\afQ}$_{\text{\fQ}}$ and {\afQ}$_{\text{\fQ}}$-{\fO} decreases, whereas increases for between {\afQ}$_{\text{\fQ}}$ and paramagnet.
(b) Phase diagram of coexisting quadrupole-octupolar order as functions of magnetic field ${\boldsymbol B}//(110)$. In this case, magnetic fields induce the increase of phase transition temperature for both cases. See the main text for details.}
\label{fig:PD-QuadOct-B}
\end{figure}

Finally, let us turn to the field evolution in the case where we assume $r_{u\phi}$ is large and positive and integrated out $\phi_u$, 
but study the interplay of $\phi_s$ and $m_u$ as we have
done at $B=0$. We thus minimize the free energy ${\cal F}^{\rm eff}_{\phi s}+ {\cal F}_{m u}+{\cal F}^{(4)}_{\rm int}$,
and supplement this with 
\be
{\cal F}^{\rm eff}_B = i B^2  \gamma_{\rm eff} (\psi_B^\pdg \phi_s^2 - \psi_B^* \phi_s^{*2}) + B^2 \tilde{r}_{s B}  |{\phi_s} |^2 \,,
\ee
where the term $\gamma_{\rm eff}$ arises from the coupling $\gamma$ in Eq.~\ref{eq:F_Bphis} upon integrating out $\phi_u$. 
Fig.~\ref{fig:PD-QuadOct-B} shows the direction dependent field evolution of multiple transitions for coexisting {\afQ} and octupolar orders. When a magnetic field is applied, the transition temperature (blue lines in Fig.~\ref{fig:PD-QuadOct-B}(a) and (b)) between paramagnet (PM) and {\afQ} with parasitic {\fQ} phase ({\afQ}$_{\text{\fQ}}$) increases due to field coupling term $\tilde{r}_{s B}$ which is 
quadratic in $\phi_s$, and since the phase is directly locked to the field direction. On the other hand, the lower transition temperature strongly depends on field direction; 
it decreases with ${\boldsymbol B}//(001)$ (red line in Fig.~\ref{fig:PD-QuadOct-B}(a)) and increases with ${\boldsymbol B}//(110)$ (red line in Fig.~\ref{fig:PD-QuadOct-B}(b)). The decrease of transition temperature with field (001) originates from the competition between the sixth order anisotropy term and field coupling terms for finite magnitudes of $\phi_s$. Thus, an anisotropic evolution of the phase diagram in a magnetic field can be also present due to {\afQ} and octupolar orders.

\section{ Discussion}

In this paper, we have formulated and studied the Landau theory of multipolar orderings in the Pr(TM)$_2$Al$_{20}$ systems, including quadrupolar and octupolar
orders. In the absence of any octupolar order, the phases of the Landau theory preserve time-reversal symmetry. In this case, examining the different quadrupolar
orders, we find that while a single thermal transition is expected in the case of {\fQ} order,
there may be multiple thermal transitions for the case of {\afQ} orders. Such a scenario involves a higher temperature transition 
from a paramagnetic phase into an {\afQ} order which breaks all
point group symmetries except $S_{4z}$, followed by a lower temperature transition into a phase where this residual Ising symmetry is broken. 
The
residual $S_{4z}$ symmetry in the intermediate phase has implications for $^{27}$Al NMR experiments which probe the induced dipole order for a `probe' magnetic
field applied along the (111) direction. For a (111) field,
there are a set of `3c' Al sites on the Frank-Kasper cage which are symmetry equivalent in the paramagnetic phase, and yield
a single NMR line.\cite{taniguchi2016nmr}
Based on symmetry, an {\afQ}-III state with $S_{4z}$ symmetry is expected to split this into four NMR lines, with a $1$:$2$ intensity ratio
(i.e., two weak and two strong).
However, the lower temperature {\afQ}-II state with broken $S_{4z}$ should exhibit six NMR lines with equal intensity. Thus, upon cooling from the 
{\afQ}-III state, which preserves $S_{4z}$ symmetry, into the low temperature {\afQ}-II state with broken $S_{4z}$ symmetry, each of the
two  original high intensity lines should split into two peaks.
Alternatively, 
the lower temperature transition may be from an intermediate {\afQ}-III state which preserves $S_{4z}$ and time reversal into a state where time-reversal
is broken by the octupolar order. In this case, the NMR should show four lines with a $1$:$2$ intensity ratio in both broken symmetry phases assuming
that the octupolar order is only weakly affected by field, but the time
reversal breaking or distinctions between {\fO} and {\afO} could be possibly detectable by $\mu$SR.\cite{u2011musr}
Further work is needed to understand the role of domains and nature of domain walls in systems with such multipolar
orders due to possible spin-lattice couplings. Clarifying the nature of these multipolar orders in the Pr(TM)$_2$Al$_{20}$ systems
would be a significant step in understanding the phase diagram and quantum critical points of such multipolar Kondo materials.

\begin{acknowledgments}
We thank S. Nakatsuji for many useful discussions, and for informing us
about their unpublished data on magnetic field dependence of the specific heat in PrV$_2$Al$_{20}$.
We also thank F. Freyer and J. Attig for their insights from an ongoing collaboration on numerical studies of 
such multipolar orders.
S.B.L. is supported by the KAIST startup and National Research Foundation Grant (NRF-2017R1A2B4008097). 
S.T. acknowledges partial funding from the DFG within CRC 1238 (project C02).
A.P. and Y.B.K. are supported by the NSERC of Canada.
This research was initiated during the 2017 ``Intertwined Orders" workshop at the 
Kavli Institute for Theoretical Physics (KITP), which is supported in part by the National Science Foundation 
under Grant No. NSF PHY-1125915. Y.B.K. also acknowledges the hospitality of the Aspen Center for Physics, 
supported in part by NSF grant PHY-1607611. 
\end{acknowledgments}



\begin{thebibliography}{46}%
\makeatletter
\providecommand \@ifxundefined [1]{%
 \@ifx{#1\undefined}
}%
\providecommand \@ifnum [1]{%
 \ifnum #1\expandafter \@firstoftwo
 \else \expandafter \@secondoftwo
 \fi
}%
\providecommand \@ifx [1]{%
 \ifx #1\expandafter \@firstoftwo
 \else \expandafter \@secondoftwo
 \fi
}%
\providecommand \natexlab [1]{#1}%
\providecommand \enquote  [1]{``#1''}%
\providecommand \bibnamefont  [1]{#1}%
\providecommand \bibfnamefont [1]{#1}%
\providecommand \citenamefont [1]{#1}%
\providecommand \href@noop [0]{\@secondoftwo}%
\providecommand \href [0]{\begingroup \@sanitize@url \@href}%
\providecommand \@href[1]{\@@startlink{#1}\@@href}%
\providecommand \@@href[1]{\endgroup#1\@@endlink}%
\providecommand \@sanitize@url [0]{\catcode `\\12\catcode `\$12\catcode
  `\&12\catcode `\#12\catcode `\^12\catcode `\_12\catcode `\%12\relax}%
\providecommand \@@startlink[1]{}%
\providecommand \@@endlink[0]{}%
\providecommand \url  [0]{\begingroup\@sanitize@url \@url }%
\providecommand \@url [1]{\endgroup\@href {#1}{\urlprefix }}%
\providecommand \urlprefix  [0]{URL }%
\providecommand \Eprint [0]{\href }%
\providecommand \doibase [0]{http://dx.doi.org/}%
\providecommand \selectlanguage [0]{\@gobble}%
\providecommand \bibinfo  [0]{\@secondoftwo}%
\providecommand \bibfield  [0]{\@secondoftwo}%
\providecommand \translation [1]{[#1]}%
\providecommand \BibitemOpen [0]{}%
\providecommand \bibitemStop [0]{}%
\providecommand \bibitemNoStop [0]{.\EOS\space}%
\providecommand \EOS [0]{\spacefactor3000\relax}%
\providecommand \BibitemShut  [1]{\csname bibitem#1\endcsname}%
\let\auto@bib@innerbib\@empty
\bibitem [{\citenamefont {Chandra}\ \emph {et~al.}(2002)\citenamefont
  {Chandra}, \citenamefont {Coleman}, \citenamefont {Mydosh},\ and\
  \citenamefont {Tripathi}}]{Chandra:2002cz}%
  \BibitemOpen
  \bibfield  {author} {\bibinfo {author} {\bibfnamefont {P}~\bibnamefont
  {Chandra}}, \bibinfo {author} {\bibfnamefont {P}~\bibnamefont {Coleman}},
  \bibinfo {author} {\bibfnamefont {J~A}\ \bibnamefont {Mydosh}}, \ and\
  \bibinfo {author} {\bibfnamefont {V}~\bibnamefont {Tripathi}},\ }\bibfield
  {title} {\enquote {\bibinfo {title} {{Hidden orbital order in the heavy
  fermion metal URu2Si2}},}\ }\href@noop {} {\bibfield  {journal} {\bibinfo
  {journal} {Nature}\ }\textbf {\bibinfo {volume} {417}},\ \bibinfo {pages}
  {831--834} (\bibinfo {year} {2002})}\BibitemShut {NoStop}%
\bibitem [{\citenamefont {Kuramoto}\ \emph {et~al.}(2009)\citenamefont
  {Kuramoto}, \citenamefont {Kusunose},\ and\ \citenamefont
  {Kiss}}]{kuramoto2009multipole}%
  \BibitemOpen
  \bibfield  {author} {\bibinfo {author} {\bibfnamefont {Yoshio}\ \bibnamefont
  {Kuramoto}}, \bibinfo {author} {\bibfnamefont {Hiroaki}\ \bibnamefont
  {Kusunose}}, \ and\ \bibinfo {author} {\bibfnamefont {Annamaria}\
  \bibnamefont {Kiss}},\ }\bibfield  {title} {\enquote {\bibinfo {title}
  {{Multipole orders and fluctuations in strongly correlated electron
  systems}},}\ }\href@noop {} {\bibfield  {journal} {\bibinfo  {journal}
  {Journal of the Physical Society of Japan}\ }\textbf {\bibinfo {volume}
  {78}},\ \bibinfo {pages} {072001} (\bibinfo {year} {2009})}\BibitemShut
  {NoStop}%
\bibitem [{\citenamefont {Kusunose}(2008)}]{kusunose2008description}%
  \BibitemOpen
  \bibfield  {author} {\bibinfo {author} {\bibfnamefont {Hiroaki}\ \bibnamefont
  {Kusunose}},\ }\bibfield  {title} {\enquote {\bibinfo {title} {{Description
  of multipole in f-electron systems}},}\ }\href@noop {} {\bibfield  {journal}
  {\bibinfo  {journal} {Journal of the Physical Society of Japan}\ }\textbf
  {\bibinfo {volume} {77}},\ \bibinfo {pages} {064710--064710} (\bibinfo {year}
  {2008})}\BibitemShut {NoStop}%
\bibitem [{\citenamefont {Santini}\ \emph {et~al.}(2009)\citenamefont
  {Santini}, \citenamefont {Carretta}, \citenamefont {Amoretti}, \citenamefont
  {Caciuffo}, \citenamefont {Magnani},\ and\ \citenamefont
  {Lander}}]{santini2009multipolar}%
  \BibitemOpen
  \bibfield  {author} {\bibinfo {author} {\bibfnamefont {Paolo}\ \bibnamefont
  {Santini}}, \bibinfo {author} {\bibfnamefont {Stefano}\ \bibnamefont
  {Carretta}}, \bibinfo {author} {\bibfnamefont {Giuseppe}\ \bibnamefont
  {Amoretti}}, \bibinfo {author} {\bibfnamefont {Roberto}\ \bibnamefont
  {Caciuffo}}, \bibinfo {author} {\bibfnamefont {Nicola}\ \bibnamefont
  {Magnani}}, \ and\ \bibinfo {author} {\bibfnamefont {Gerard~H}\ \bibnamefont
  {Lander}},\ }\bibfield  {title} {\enquote {\bibinfo {title} {{Multipolar
  interactions in f-electron systems: The paradigm of actinide dioxides}},}\
  }\href@noop {} {\bibfield  {journal} {\bibinfo  {journal} {Reviews of Modern
  Physics}\ }\textbf {\bibinfo {volume} {81}},\ \bibinfo {pages} {807}
  (\bibinfo {year} {2009})}\BibitemShut {NoStop}%
\bibitem [{\citenamefont {Shiina}\ \emph {et~al.}(1997)\citenamefont {Shiina},
  \citenamefont {Shiba},\ and\ \citenamefont {Thalmeier}}]{shiina1997magnetic}%
  \BibitemOpen
  \bibfield  {author} {\bibinfo {author} {\bibfnamefont {Ryousuke}\
  \bibnamefont {Shiina}}, \bibinfo {author} {\bibfnamefont {Hiroyuki}\
  \bibnamefont {Shiba}}, \ and\ \bibinfo {author} {\bibfnamefont {Peter}\
  \bibnamefont {Thalmeier}},\ }\bibfield  {title} {\enquote {\bibinfo {title}
  {{Magnetic-field effects on quadrupolar ordering in a $\Gamma$ 8-quartet
  system CeB$_6$}},}\ }\href@noop {} {\bibfield  {journal} {\bibinfo  {journal}
  {Journal of the Physical Society of Japan}\ }\textbf {\bibinfo {volume}
  {66}},\ \bibinfo {pages} {1741--1755} (\bibinfo {year} {1997})}\BibitemShut
  {NoStop}%
\bibitem [{\citenamefont {Kitagawa}\ \emph {et~al.}(1996)\citenamefont
  {Kitagawa}, \citenamefont {Takeda},\ and\ \citenamefont
  {Ishikawa}}]{kitagawa1996possible}%
  \BibitemOpen
  \bibfield  {author} {\bibinfo {author} {\bibfnamefont {Jiro}\ \bibnamefont
  {Kitagawa}}, \bibinfo {author} {\bibfnamefont {Naoya}\ \bibnamefont
  {Takeda}}, \ and\ \bibinfo {author} {\bibfnamefont {Masayasu}\ \bibnamefont
  {Ishikawa}},\ }\bibfield  {title} {\enquote {\bibinfo {title} {{Possible
  quadrupolar ordering in a Kondo-lattice compound Ce$_3$Pd$_{20}$Ge$_6$}},}\
  }\href@noop {} {\bibfield  {journal} {\bibinfo  {journal} {Phys. Rev. B}\
  }\textbf {\bibinfo {volume} {53}},\ \bibinfo {pages} {5101} (\bibinfo {year}
  {1996})}\BibitemShut {NoStop}%
\bibitem [{\citenamefont {Caciuffo}\ \emph {et~al.}(2003)\citenamefont
  {Caciuffo}, \citenamefont {Paix{\~a}o}, \citenamefont {Detlefs},
  \citenamefont {Longfield}, \citenamefont {Santini}, \citenamefont
  {Bernhoeft}, \citenamefont {Rebizant},\ and\ \citenamefont
  {Lander}}]{caciuffo2003multipolar}%
  \BibitemOpen
  \bibfield  {author} {\bibinfo {author} {\bibfnamefont {R.}~\bibnamefont
  {Caciuffo}}, \bibinfo {author} {\bibfnamefont {J.A.}\ \bibnamefont
  {Paix{\~a}o}}, \bibinfo {author} {\bibfnamefont {C.}~\bibnamefont {Detlefs}},
  \bibinfo {author} {\bibfnamefont {M.J.}\ \bibnamefont {Longfield}}, \bibinfo
  {author} {\bibfnamefont {P.}~\bibnamefont {Santini}}, \bibinfo {author}
  {\bibfnamefont {N.}~\bibnamefont {Bernhoeft}}, \bibinfo {author}
  {\bibfnamefont {J.}~\bibnamefont {Rebizant}}, \ and\ \bibinfo {author}
  {\bibfnamefont {G.H.}\ \bibnamefont {Lander}},\ }\bibfield  {title} {\enquote
  {\bibinfo {title} {{Multipolar ordering in NpO$_2$ below 25 K}},}\
  }\href@noop {} {\bibfield  {journal} {\bibinfo  {journal} {Journal of
  Physics: Condensed Matter}\ }\textbf {\bibinfo {volume} {15}},\ \bibinfo
  {pages} {S2287} (\bibinfo {year} {2003})}\BibitemShut {NoStop}%
\bibitem [{\citenamefont {Morin}\ \emph {et~al.}(1982)\citenamefont {Morin},
  \citenamefont {Schmitt},\ and\ \citenamefont
  {De~Lacheisserie}}]{morin1982magnetic}%
  \BibitemOpen
  \bibfield  {author} {\bibinfo {author} {\bibfnamefont {P.}~\bibnamefont
  {Morin}}, \bibinfo {author} {\bibfnamefont {D.}~\bibnamefont {Schmitt}}, \
  and\ \bibinfo {author} {\bibfnamefont {E.~Du~Tremolet}\ \bibnamefont
  {De~Lacheisserie}},\ }\bibfield  {title} {\enquote {\bibinfo {title}
  {{Magnetic and quadrupolar properties of PrPb$_3$}},}\ }\href@noop {}
  {\bibfield  {journal} {\bibinfo  {journal} {Journal of Magnetism and Magnetic
  Materials}\ }\textbf {\bibinfo {volume} {30}},\ \bibinfo {pages} {257}
  (\bibinfo {year} {1982})}\BibitemShut {NoStop}%
\bibitem [{\citenamefont {Lee}\ \emph {et~al.}(2015)\citenamefont {Lee},
  \citenamefont {Paramekanti},\ and\ \citenamefont {Kim}}]{lee2015optical}%
  \BibitemOpen
  \bibfield  {author} {\bibinfo {author} {\bibfnamefont {SungBin}\ \bibnamefont
  {Lee}}, \bibinfo {author} {\bibfnamefont {Arun}\ \bibnamefont {Paramekanti}},
  \ and\ \bibinfo {author} {\bibfnamefont {Yong~Baek}\ \bibnamefont {Kim}},\
  }\bibfield  {title} {\enquote {\bibinfo {title} {{Optical gyrotropy in
  quadrupolar Kondo systems}},}\ }\href@noop {} {\bibfield  {journal} {\bibinfo
   {journal} {Phys. Rev. B}\ }\textbf {\bibinfo {volume} {91}},\ \bibinfo
  {pages} {041104} (\bibinfo {year} {2015})}\BibitemShut {NoStop}%
\bibitem [{\citenamefont {Suzuki}\ \emph {et~al.}(2005)\citenamefont {Suzuki},
  \citenamefont {S.~Suzuki}, \citenamefont {Kitazawa}, \citenamefont {Kido},
  \citenamefont {Ueno}, \citenamefont {Yamaguchi}, \citenamefont {Nemoto},\
  and\ \citenamefont {Goto}}]{suzuki2005quadrupolar}%
  \BibitemOpen
  \bibfield  {author} {\bibinfo {author} {\bibfnamefont {Osamu}\ \bibnamefont
  {Suzuki}}, \bibinfo {author} {\bibfnamefont {Hiroyuki}\ \bibnamefont
  {S.~Suzuki}}, \bibinfo {author} {\bibfnamefont {Hideaki}\ \bibnamefont
  {Kitazawa}}, \bibinfo {author} {\bibfnamefont {Giyuu}\ \bibnamefont {Kido}},
  \bibinfo {author} {\bibfnamefont {Takafumi}\ \bibnamefont {Ueno}}, \bibinfo
  {author} {\bibfnamefont {Takashi}\ \bibnamefont {Yamaguchi}}, \bibinfo
  {author} {\bibfnamefont {Yuichi}\ \bibnamefont {Nemoto}}, \ and\ \bibinfo
  {author} {\bibfnamefont {Terutaka}\ \bibnamefont {Goto}},\ }\bibfield
  {title} {\enquote {\bibinfo {title} {{Quadrupolar Kondo effect in non-Kramers
  doublet system PrInAg$_2$}},}\ }\href@noop {} {\bibfield  {journal} {\bibinfo
   {journal} {J. Phys. Soc. Jpn.}\ }\textbf {\bibinfo {volume} {75}},\ \bibinfo
  {pages} {013704} (\bibinfo {year} {2005})}\BibitemShut {NoStop}%
\bibitem [{\citenamefont {Sakai}\ and\ \citenamefont
  {Nakatsuji}(2012)}]{sakai2012thermal}%
  \BibitemOpen
  \bibfield  {author} {\bibinfo {author} {\bibfnamefont {Akito}\ \bibnamefont
  {Sakai}}\ and\ \bibinfo {author} {\bibfnamefont {Satoru}\ \bibnamefont
  {Nakatsuji}},\ }\bibfield  {title} {\enquote {\bibinfo {title} {{Thermal
  properties of the nonmagnetic cubic $\Gamma$3 Kondo lattice systems
  PrTr$_2$Al$_{20}$ (Tr=Ti, V)}},}\ }in\ \href@noop {} {\emph {\bibinfo
  {booktitle} {Journal of Physics: Conference Series}}},\ Vol.\ \bibinfo
  {volume} {391}\ (\bibinfo {organization} {IOP Publishing},\ \bibinfo {year}
  {2012})\ p.\ \bibinfo {pages} {012058}\BibitemShut {NoStop}%
\bibitem [{\citenamefont {Sato}\ \emph {et~al.}(2012)\citenamefont {Sato},
  \citenamefont {Ibuka}, \citenamefont {Nambu}, \citenamefont {Yamazaki},
  \citenamefont {Hong}, \citenamefont {Sakai},\ and\ \citenamefont
  {Nakatsuji}}]{sato2012ferroquadrupolar}%
  \BibitemOpen
  \bibfield  {author} {\bibinfo {author} {\bibfnamefont {Taku~J.}\ \bibnamefont
  {Sato}}, \bibinfo {author} {\bibfnamefont {Soshi}\ \bibnamefont {Ibuka}},
  \bibinfo {author} {\bibfnamefont {Yusuke}\ \bibnamefont {Nambu}}, \bibinfo
  {author} {\bibfnamefont {Teruo}\ \bibnamefont {Yamazaki}}, \bibinfo {author}
  {\bibfnamefont {Tao}\ \bibnamefont {Hong}}, \bibinfo {author} {\bibfnamefont
  {Akito}\ \bibnamefont {Sakai}}, \ and\ \bibinfo {author} {\bibfnamefont
  {Satoru}\ \bibnamefont {Nakatsuji}},\ }\bibfield  {title} {\enquote {\bibinfo
  {title} {{Ferroquadrupolar ordering in PrTi$_2$Al$_{20}$}},}\ }\href@noop {}
  {\bibfield  {journal} {\bibinfo  {journal} {Phys. Rev. B}\ }\textbf {\bibinfo
  {volume} {86}},\ \bibinfo {pages} {184419} (\bibinfo {year}
  {2012})}\BibitemShut {NoStop}%
\bibitem [{\citenamefont {Onimaru}\ and\ \citenamefont
  {Kusunose}(2016)}]{onimaru2016exotic}%
  \BibitemOpen
  \bibfield  {author} {\bibinfo {author} {\bibfnamefont {Takahiro}\
  \bibnamefont {Onimaru}}\ and\ \bibinfo {author} {\bibfnamefont {Hiroaki}\
  \bibnamefont {Kusunose}},\ }\bibfield  {title} {\enquote {\bibinfo {title}
  {{Exotic Quadrupolar Phenomena in Non-Kramers Doublet Systems? The Cases of
  PrT$_2$Zn$_{20}$ (T=Ir, Rh) and PrT$_2$Al$_{20}$ (T=V, Ti)?}}}\ }\href@noop
  {} {\bibfield  {journal} {\bibinfo  {journal} {J. Phys. Soc. Jpn.}\ }\textbf
  {\bibinfo {volume} {85}},\ \bibinfo {pages} {082002} (\bibinfo {year}
  {2016})}\BibitemShut {NoStop}%
\bibitem [{\citenamefont {Onimaru}\ \emph {et~al.}(2011)\citenamefont
  {Onimaru}, \citenamefont {Matsumoto}, \citenamefont {Inoue}, \citenamefont
  {Umeo}, \citenamefont {Sakakibara}, \citenamefont {Karaki}, \citenamefont
  {Kubota},\ and\ \citenamefont
  {Takabatake}}]{onimaru2011antiferroquadrupolar}%
  \BibitemOpen
  \bibfield  {author} {\bibinfo {author} {\bibfnamefont {T.}~\bibnamefont
  {Onimaru}}, \bibinfo {author} {\bibfnamefont {K.T.}\ \bibnamefont
  {Matsumoto}}, \bibinfo {author} {\bibfnamefont {Y.F.}\ \bibnamefont {Inoue}},
  \bibinfo {author} {\bibfnamefont {K.}~\bibnamefont {Umeo}}, \bibinfo {author}
  {\bibfnamefont {T.}~\bibnamefont {Sakakibara}}, \bibinfo {author}
  {\bibfnamefont {Y.}~\bibnamefont {Karaki}}, \bibinfo {author} {\bibfnamefont
  {M.}~\bibnamefont {Kubota}}, \ and\ \bibinfo {author} {\bibfnamefont
  {T.}~\bibnamefont {Takabatake}},\ }\bibfield  {title} {\enquote {\bibinfo
  {title} {{Antiferroquadrupolar ordering in a Pr-based superconductor
  PrIr$_2$Zn$_20$}},}\ }\href@noop {} {\bibfield  {journal} {\bibinfo
  {journal} {Phys. Rev. Lett.}\ }\textbf {\bibinfo {volume} {106}},\ \bibinfo
  {pages} {177001} (\bibinfo {year} {2011})}\BibitemShut {NoStop}%
\bibitem [{\citenamefont {Tsujimoto}\ \emph {et~al.}(2015)\citenamefont
  {Tsujimoto}, \citenamefont {Matsumoto},\ and\ \citenamefont
  {Nakatsuji}}]{tsujimoto2015anomalous}%
  \BibitemOpen
  \bibfield  {author} {\bibinfo {author} {\bibfnamefont {Masaki}\ \bibnamefont
  {Tsujimoto}}, \bibinfo {author} {\bibfnamefont {Yosuke}\ \bibnamefont
  {Matsumoto}}, \ and\ \bibinfo {author} {\bibfnamefont {Satoru}\ \bibnamefont
  {Nakatsuji}},\ }\bibfield  {title} {\enquote {\bibinfo {title} {{Anomalous
  specific heat behaviour in the quadrupolar Kondo system PrV$_2$Al$_{20}$}},}\
  }in\ \href@noop {} {\emph {\bibinfo {booktitle} {Journal of Physics:
  Conference Series}}},\ Vol.\ \bibinfo {volume} {592}\ (\bibinfo
  {organization} {IOP Publishing},\ \bibinfo {year} {2015})\ p.\ \bibinfo
  {pages} {012023}\BibitemShut {NoStop}%
\bibitem [{\citenamefont {Onimaru}\ \emph {et~al.}(2012)\citenamefont
  {Onimaru}, \citenamefont {Nagasawa}, \citenamefont {Matsumoto}, \citenamefont
  {Wakiya}, \citenamefont {Umeo}, \citenamefont {Kittaka}, \citenamefont
  {Sakakibara}, \citenamefont {Matsushita},\ and\ \citenamefont
  {Takabatake}}]{onimaru2012simultaneous}%
  \BibitemOpen
  \bibfield  {author} {\bibinfo {author} {\bibfnamefont {T.}~\bibnamefont
  {Onimaru}}, \bibinfo {author} {\bibfnamefont {N.}~\bibnamefont {Nagasawa}},
  \bibinfo {author} {\bibfnamefont {K.T.}\ \bibnamefont {Matsumoto}}, \bibinfo
  {author} {\bibfnamefont {K.}~\bibnamefont {Wakiya}}, \bibinfo {author}
  {\bibfnamefont {K.}~\bibnamefont {Umeo}}, \bibinfo {author} {\bibfnamefont
  {S.}~\bibnamefont {Kittaka}}, \bibinfo {author} {\bibfnamefont
  {T.}~\bibnamefont {Sakakibara}}, \bibinfo {author} {\bibfnamefont
  {Y.}~\bibnamefont {Matsushita}}, \ and\ \bibinfo {author} {\bibfnamefont
  {T.}~\bibnamefont {Takabatake}},\ }\bibfield  {title} {\enquote {\bibinfo
  {title} {{Simultaneous superconducting and antiferroquadrupolar transitions
  in PrRh$_2$Zn$_{20}$}},}\ }\href@noop {} {\bibfield  {journal} {\bibinfo
  {journal} {Phys. Rev. B}\ }\textbf {\bibinfo {volume} {86}},\ \bibinfo
  {pages} {184426} (\bibinfo {year} {2012})}\BibitemShut {NoStop}%
\bibitem [{\citenamefont {Onimaru}\ \emph {et~al.}(2010)\citenamefont
  {Onimaru}, \citenamefont {T.~Matsumoto}, \citenamefont {F.~Inoue},
  \citenamefont {Umeo}, \citenamefont {Saiga}, \citenamefont {Matsushita},
  \citenamefont {Tamura}, \citenamefont {Nishimoto}, \citenamefont {Ishii},
  \citenamefont {Suzuki} \emph {et~al.}}]{onimaru2010superconductivity}%
  \BibitemOpen
  \bibfield  {author} {\bibinfo {author} {\bibfnamefont {Takahiro}\
  \bibnamefont {Onimaru}}, \bibinfo {author} {\bibfnamefont {Keisuke}\
  \bibnamefont {T.~Matsumoto}}, \bibinfo {author} {\bibfnamefont {Yukihiro}\
  \bibnamefont {F.~Inoue}}, \bibinfo {author} {\bibfnamefont {Kazunori}\
  \bibnamefont {Umeo}}, \bibinfo {author} {\bibfnamefont {Yuta}\ \bibnamefont
  {Saiga}}, \bibinfo {author} {\bibfnamefont {Yoshitaka}\ \bibnamefont
  {Matsushita}}, \bibinfo {author} {\bibfnamefont {Ryuji}\ \bibnamefont
  {Tamura}}, \bibinfo {author} {\bibfnamefont {Kazue}\ \bibnamefont
  {Nishimoto}}, \bibinfo {author} {\bibfnamefont {Isao}\ \bibnamefont {Ishii}},
  \bibinfo {author} {\bibfnamefont {Takashi}\ \bibnamefont {Suzuki}},  \emph
  {et~al.},\ }\bibfield  {title} {\enquote {\bibinfo {title}
  {{Superconductivity and structural phase transitions in caged compounds
  RT$_2$Zn$_{20}$ (R=La, Pr, T=Ru, Ir)}},}\ }\href@noop {} {\bibfield
  {journal} {\bibinfo  {journal} {J. Phys. Soc. Jpn.}\ }\textbf {\bibinfo
  {volume} {79}},\ \bibinfo {pages} {033704} (\bibinfo {year}
  {2010})}\BibitemShut {NoStop}%
\bibitem [{\citenamefont {Sakai}\ \emph {et~al.}(2012)\citenamefont {Sakai},
  \citenamefont {Kuga},\ and\ \citenamefont
  {Nakatsuji}}]{sakai2012superconductivity}%
  \BibitemOpen
  \bibfield  {author} {\bibinfo {author} {\bibfnamefont {Akito}\ \bibnamefont
  {Sakai}}, \bibinfo {author} {\bibfnamefont {Kentaro}\ \bibnamefont {Kuga}}, \
  and\ \bibinfo {author} {\bibfnamefont {Satoru}\ \bibnamefont {Nakatsuji}},\
  }\bibfield  {title} {\enquote {\bibinfo {title} {{Superconductivity in the
  ferroquadrupolar state in the quadrupolar Kondo lattice
  PrTi$_2$Al$_{20}$}},}\ }\href@noop {} {\bibfield  {journal} {\bibinfo
  {journal} {J. Phys. Soc. Jpn.}\ }\textbf {\bibinfo {volume} {81}},\ \bibinfo
  {pages} {083702} (\bibinfo {year} {2012})}\BibitemShut {NoStop}%
\bibitem [{\citenamefont {Matsubayashi}\ \emph {et~al.}(2012)\citenamefont
  {Matsubayashi}, \citenamefont {Tanaka}, \citenamefont {Sakai}, \citenamefont
  {Nakatsuji}, \citenamefont {Kubo},\ and\ \citenamefont
  {Uwatoko}}]{matsubayashi2012pressure}%
  \BibitemOpen
  \bibfield  {author} {\bibinfo {author} {\bibfnamefont {K.}~\bibnamefont
  {Matsubayashi}}, \bibinfo {author} {\bibfnamefont {T.}~\bibnamefont
  {Tanaka}}, \bibinfo {author} {\bibfnamefont {A.}~\bibnamefont {Sakai}},
  \bibinfo {author} {\bibfnamefont {S.}~\bibnamefont {Nakatsuji}}, \bibinfo
  {author} {\bibfnamefont {Y.}~\bibnamefont {Kubo}}, \ and\ \bibinfo {author}
  {\bibfnamefont {Y.}~\bibnamefont {Uwatoko}},\ }\bibfield  {title} {\enquote
  {\bibinfo {title} {{Pressure-induced heavy fermion superconductivity in the
  nonmagnetic quadrupolar system PrTi$_2$Al$_{20}$}},}\ }\href@noop {}
  {\bibfield  {journal} {\bibinfo  {journal} {Phys. Rev. Lett.}\ }\textbf
  {\bibinfo {volume} {109}},\ \bibinfo {pages} {187004} (\bibinfo {year}
  {2012})}\BibitemShut {NoStop}%
\bibitem [{\citenamefont {Matsubayashi}\ \emph {et~al.}(2014)\citenamefont
  {Matsubayashi}, \citenamefont {Tanaka}, \citenamefont {Suzuki}, \citenamefont
  {Sakai}, \citenamefont {Nakatsuji}, \citenamefont {Kitagawa}, \citenamefont
  {Kubo},\ and\ \citenamefont {Uwatoko}}]{matsubayashi2014heavy}%
  \BibitemOpen
  \bibfield  {author} {\bibinfo {author} {\bibfnamefont {Kazuyuki}\
  \bibnamefont {Matsubayashi}}, \bibinfo {author} {\bibfnamefont {Toshiki}\
  \bibnamefont {Tanaka}}, \bibinfo {author} {\bibfnamefont {Junichirou}\
  \bibnamefont {Suzuki}}, \bibinfo {author} {\bibfnamefont {Akito}\
  \bibnamefont {Sakai}}, \bibinfo {author} {\bibfnamefont {Satoru}\
  \bibnamefont {Nakatsuji}}, \bibinfo {author} {\bibfnamefont {Kentaro}\
  \bibnamefont {Kitagawa}}, \bibinfo {author} {\bibfnamefont {Yasunori}\
  \bibnamefont {Kubo}}, \ and\ \bibinfo {author} {\bibfnamefont {Yoshiya}\
  \bibnamefont {Uwatoko}},\ }\bibfield  {title} {\enquote {\bibinfo {title}
  {{Heavy Fermion Superconductivity under Pressure in the Quadrupole System
  PrTi$_2$Al$_{20}$}},}\ }in\ \href@noop {} {\emph {\bibinfo {booktitle}
  {Proceedings of the International Conference on Strongly Correlated Electron
  Systems (SCES2013)}}}\ (\bibinfo {year} {2014})\ p.\ \bibinfo {pages}
  {011077}\BibitemShut {NoStop}%
\bibitem [{\citenamefont {Tsujimoto}\ \emph {et~al.}(2014)\citenamefont
  {Tsujimoto}, \citenamefont {Matsumoto}, \citenamefont {Tomita}, \citenamefont
  {Sakai},\ and\ \citenamefont {Nakatsuji}}]{tsujimoto2014heavy}%
  \BibitemOpen
  \bibfield  {author} {\bibinfo {author} {\bibfnamefont {Masaki}\ \bibnamefont
  {Tsujimoto}}, \bibinfo {author} {\bibfnamefont {Yosuke}\ \bibnamefont
  {Matsumoto}}, \bibinfo {author} {\bibfnamefont {Takahiro}\ \bibnamefont
  {Tomita}}, \bibinfo {author} {\bibfnamefont {Akito}\ \bibnamefont {Sakai}}, \
  and\ \bibinfo {author} {\bibfnamefont {Satoru}\ \bibnamefont {Nakatsuji}},\
  }\bibfield  {title} {\enquote {\bibinfo {title} {{Heavy-fermion
  superconductivity in the quadrupole ordered state of PrV$_2$Al$_{20}$}},}\
  }\href@noop {} {\bibfield  {journal} {\bibinfo  {journal} {Phys. Rev. Lett.}\
  }\textbf {\bibinfo {volume} {113}},\ \bibinfo {pages} {267001} (\bibinfo
  {year} {2014})}\BibitemShut {NoStop}%
\bibitem [{\citenamefont {Iwasa}\ \emph {et~al.}(2017)\citenamefont {Iwasa},
  \citenamefont {Matsumoto}, \citenamefont {Onimaru}, \citenamefont
  {Takabatake}, \citenamefont {Mignot},\ and\ \citenamefont
  {Gukasov}}]{iwasa2017evidence}%
  \BibitemOpen
  \bibfield  {author} {\bibinfo {author} {\bibfnamefont {Kazuaki}\ \bibnamefont
  {Iwasa}}, \bibinfo {author} {\bibfnamefont {Keisuke~T.}\ \bibnamefont
  {Matsumoto}}, \bibinfo {author} {\bibfnamefont {Takahiro}\ \bibnamefont
  {Onimaru}}, \bibinfo {author} {\bibfnamefont {Toshiro}\ \bibnamefont
  {Takabatake}}, \bibinfo {author} {\bibfnamefont {Jean-Michel}\ \bibnamefont
  {Mignot}}, \ and\ \bibinfo {author} {\bibfnamefont {Arsen}\ \bibnamefont
  {Gukasov}},\ }\bibfield  {title} {\enquote {\bibinfo {title} {{Evidence for
  antiferromagnetic-type ordering of f-electron multipoles in
  PrIr$_2$Zn$_{20}$}},}\ }\href@noop {} {\bibfield  {journal} {\bibinfo
  {journal} {Phys. Rev. B}\ }\textbf {\bibinfo {volume} {95}},\ \bibinfo
  {pages} {155106} (\bibinfo {year} {2017})}\BibitemShut {NoStop}%
\bibitem [{\citenamefont {U.~Ito}\ \emph {et~al.}(2011)\citenamefont {U.~Ito},
  \citenamefont {Higemoto}, \citenamefont {Ninomiya}, \citenamefont {Luetkens},
  \citenamefont {Baines}, \citenamefont {Sakai},\ and\ \citenamefont
  {Nakatsuji}}]{u2011musr}%
  \BibitemOpen
  \bibfield  {author} {\bibinfo {author} {\bibfnamefont {Takashi}\ \bibnamefont
  {U.~Ito}}, \bibinfo {author} {\bibfnamefont {Wataru}\ \bibnamefont
  {Higemoto}}, \bibinfo {author} {\bibfnamefont {Kazuhiko}\ \bibnamefont
  {Ninomiya}}, \bibinfo {author} {\bibfnamefont {Hubertus}\ \bibnamefont
  {Luetkens}}, \bibinfo {author} {\bibfnamefont {Christopher}\ \bibnamefont
  {Baines}}, \bibinfo {author} {\bibfnamefont {Akito}\ \bibnamefont {Sakai}}, \
  and\ \bibinfo {author} {\bibfnamefont {Satoru}\ \bibnamefont {Nakatsuji}},\
  }\bibfield  {title} {\enquote {\bibinfo {title} {{$\mu$SR Evidence of
  Nonmagnetic Order and 141Pr Hyperfine-Enhanced Nuclear Magnetism in the Cubic
  $\Gamma$3 Ground Doublet System PrTi2Al20}},}\ }\href@noop {} {\bibfield
  {journal} {\bibinfo  {journal} {Journal of the Physical Society of Japan}\
  }\textbf {\bibinfo {volume} {80}},\ \bibinfo {pages} {113703} (\bibinfo
  {year} {2011})}\BibitemShut {NoStop}%
\bibitem [{\citenamefont {Cox}\ and\ \citenamefont
  {Zawadowski}(1999)}]{cox1999exotic}%
  \BibitemOpen
  \bibfield  {author} {\bibinfo {author} {\bibfnamefont {D.L.}\ \bibnamefont
  {Cox}}\ and\ \bibinfo {author} {\bibfnamefont {Alfred}\ \bibnamefont
  {Zawadowski}},\ }\href@noop {} {\emph {\bibinfo {title} {{Exotic Kondo
  Effects in Metals: Magnetic Ions in a Crystalline Electric Field and
  Tunelling Centres}}}}\ (\bibinfo  {publisher} {CRC Press},\ \bibinfo {year}
  {1999})\BibitemShut {NoStop}%
\bibitem [{\citenamefont {Cox}(1987)}]{cox1987quadrupolar}%
  \BibitemOpen
  \bibfield  {author} {\bibinfo {author} {\bibfnamefont {D.L.}\ \bibnamefont
  {Cox}},\ }\bibfield  {title} {\enquote {\bibinfo {title} {{Quadrupolar Kondo
  effect in uranium heavy-electron materials?}}}\ }\href@noop {} {\bibfield
  {journal} {\bibinfo  {journal} {Phys. Rev. Lett.}\ }\textbf {\bibinfo
  {volume} {59}},\ \bibinfo {pages} {1240} (\bibinfo {year}
  {1987})}\BibitemShut {NoStop}%
\bibitem [{\citenamefont {Si}\ and\ \citenamefont
  {Steglich}(2010)}]{si2010heavy}%
  \BibitemOpen
  \bibfield  {author} {\bibinfo {author} {\bibfnamefont {Qimiao}\ \bibnamefont
  {Si}}\ and\ \bibinfo {author} {\bibfnamefont {Frank}\ \bibnamefont
  {Steglich}},\ }\bibfield  {title} {\enquote {\bibinfo {title} {{Heavy
  fermions and quantum phase transitions}},}\ }\href@noop {} {\bibfield
  {journal} {\bibinfo  {journal} {Science}\ }\textbf {\bibinfo {volume}
  {329}},\ \bibinfo {pages} {1161} (\bibinfo {year} {2010})}\BibitemShut
  {NoStop}%
\bibitem [{\citenamefont {Stewart}(1984)}]{stewart1984heavy}%
  \BibitemOpen
  \bibfield  {author} {\bibinfo {author} {\bibfnamefont {See~G.R.}\
  \bibnamefont {Stewart}},\ }\bibfield  {title} {\enquote {\bibinfo {title}
  {{Heavy-fermion systems}},}\ }\href@noop {} {\bibfield  {journal} {\bibinfo
  {journal} {Rev. Mod. Phys.}\ }\textbf {\bibinfo {volume} {56}},\ \bibinfo
  {pages} {755} (\bibinfo {year} {1984})}\BibitemShut {NoStop}%
\bibitem [{\citenamefont {Gegenwart}\ \emph {et~al.}(2008)\citenamefont
  {Gegenwart}, \citenamefont {Si},\ and\ \citenamefont
  {Steglich}}]{gegenwart2008quantum}%
  \BibitemOpen
  \bibfield  {author} {\bibinfo {author} {\bibfnamefont {Philipp}\ \bibnamefont
  {Gegenwart}}, \bibinfo {author} {\bibfnamefont {Qimiao}\ \bibnamefont {Si}},
  \ and\ \bibinfo {author} {\bibfnamefont {Frank}\ \bibnamefont {Steglich}},\
  }\bibfield  {title} {\enquote {\bibinfo {title} {{Quantum criticality in
  heavy-fermion metals}},}\ }\href@noop {} {\bibfield  {journal} {\bibinfo
  {journal} {Nature Physics}\ }\textbf {\bibinfo {volume} {4}},\ \bibinfo
  {pages} {186} (\bibinfo {year} {2008})}\BibitemShut {NoStop}%
\bibitem [{\citenamefont {Fisk}\ \emph {et~al.}(1995)\citenamefont {Fisk},
  \citenamefont {Sarrao}, \citenamefont {Smith},\ and\ \citenamefont
  {Thompson}}]{fisk1995physics}%
  \BibitemOpen
  \bibfield  {author} {\bibinfo {author} {\bibfnamefont {Z.}~\bibnamefont
  {Fisk}}, \bibinfo {author} {\bibfnamefont {J.L.}\ \bibnamefont {Sarrao}},
  \bibinfo {author} {\bibfnamefont {J.L.}\ \bibnamefont {Smith}}, \ and\
  \bibinfo {author} {\bibfnamefont {J.D.}\ \bibnamefont {Thompson}},\
  }\bibfield  {title} {\enquote {\bibinfo {title} {{The physics and chemistry
  of heavy fermions}},}\ }\href@noop {} {\bibfield  {journal} {\bibinfo
  {journal} {Proceedings of the National Academy of Sciences}\ }\textbf
  {\bibinfo {volume} {92}},\ \bibinfo {pages} {6663} (\bibinfo {year}
  {1995})}\BibitemShut {NoStop}%
\bibitem [{\citenamefont {Sakai}\ and\ \citenamefont
  {Nakatsuji}(2011)}]{sakai2011kondo}%
  \BibitemOpen
  \bibfield  {author} {\bibinfo {author} {\bibfnamefont {Akito}\ \bibnamefont
  {Sakai}}\ and\ \bibinfo {author} {\bibfnamefont {Satoru}\ \bibnamefont
  {Nakatsuji}},\ }\bibfield  {title} {\enquote {\bibinfo {title} {{Kondo
  Effects and Multipolar Order in the Cubic PrTr$_2$Al$_{20}$ (Tr=Ti, V)}},}\
  }\href@noop {} {\bibfield  {journal} {\bibinfo  {journal} {J. Phys. Soc.
  Jpn.}\ }\textbf {\bibinfo {volume} {80}},\ \bibinfo {pages} {063701}
  (\bibinfo {year} {2011})}\BibitemShut {NoStop}%
\bibitem [{\citenamefont {Koseki}\ \emph {et~al.}(2011)\citenamefont {Koseki},
  \citenamefont {Nakanishi}, \citenamefont {Deto}, \citenamefont {Koseki},
  \citenamefont {Kashiwazaki}, \citenamefont {Shichinomiya}, \citenamefont
  {Nakamura}, \citenamefont {Yoshizawa}, \citenamefont {Sakai},\ and\
  \citenamefont {Nakatsuji}}]{koseki2011ultrasonic}%
  \BibitemOpen
  \bibfield  {author} {\bibinfo {author} {\bibfnamefont {Minoru}\ \bibnamefont
  {Koseki}}, \bibinfo {author} {\bibfnamefont {Yoshiki}\ \bibnamefont
  {Nakanishi}}, \bibinfo {author} {\bibfnamefont {Kazuhisa}\ \bibnamefont
  {Deto}}, \bibinfo {author} {\bibfnamefont {Gen}\ \bibnamefont {Koseki}},
  \bibinfo {author} {\bibfnamefont {Reiko}\ \bibnamefont {Kashiwazaki}},
  \bibinfo {author} {\bibfnamefont {Fumitaka}\ \bibnamefont {Shichinomiya}},
  \bibinfo {author} {\bibfnamefont {Mitsuteru}\ \bibnamefont {Nakamura}},
  \bibinfo {author} {\bibfnamefont {Masahito}\ \bibnamefont {Yoshizawa}},
  \bibinfo {author} {\bibfnamefont {Akihito}\ \bibnamefont {Sakai}}, \ and\
  \bibinfo {author} {\bibfnamefont {Satoru}\ \bibnamefont {Nakatsuji}},\
  }\bibfield  {title} {\enquote {\bibinfo {title} {{Ultrasonic investigation on
  a cage structure compound PrTi$_2$Al$_{20}$}},}\ }\href@noop {} {\bibfield
  {journal} {\bibinfo  {journal} {J. Phys. Soc. Jpn.}\ }\textbf {\bibinfo
  {volume} {80}},\ \bibinfo {pages} {SA049} (\bibinfo {year}
  {2011})}\BibitemShut {NoStop}%
\bibitem [{\citenamefont {Matsumoto}\ \emph {et~al.}(2016)\citenamefont
  {Matsumoto}, \citenamefont {Tsujimoto}, \citenamefont {Tomita}, \citenamefont
  {Sakai},\ and\ \citenamefont {Nakatsuji}}]{matsumoto2016strong}%
  \BibitemOpen
  \bibfield  {author} {\bibinfo {author} {\bibfnamefont {Yosuke}\ \bibnamefont
  {Matsumoto}}, \bibinfo {author} {\bibfnamefont {Masaki}\ \bibnamefont
  {Tsujimoto}}, \bibinfo {author} {\bibfnamefont {Takahiro}\ \bibnamefont
  {Tomita}}, \bibinfo {author} {\bibfnamefont {Akito}\ \bibnamefont {Sakai}}, \
  and\ \bibinfo {author} {\bibfnamefont {Satoru}\ \bibnamefont {Nakatsuji}},\
  }\bibfield  {title} {\enquote {\bibinfo {title} {{Strong orbital fluctuations
  in multipolar ordered states of PrV2Al20}},}\ }\href@noop {} {\bibfield
  {journal} {\bibinfo  {journal} {Journal of Magnetism and Magnetic Materials}\
  }\textbf {\bibinfo {volume} {400}},\ \bibinfo {pages} {66--69} (\bibinfo
  {year} {2016})}\BibitemShut {NoStop}%
\bibitem [{\citenamefont {Tokunaga}\ \emph {et~al.}(2013)\citenamefont
  {Tokunaga}, \citenamefont {Sakai}, \citenamefont {Kambe}, \citenamefont
  {Sakai}, \citenamefont {Nakatsuji},\ and\ \citenamefont
  {Harima}}]{tokunaga2013magnetic}%
  \BibitemOpen
  \bibfield  {author} {\bibinfo {author} {\bibfnamefont {Yo}~\bibnamefont
  {Tokunaga}}, \bibinfo {author} {\bibfnamefont {Hironori}\ \bibnamefont
  {Sakai}}, \bibinfo {author} {\bibfnamefont {Shinsaku}\ \bibnamefont {Kambe}},
  \bibinfo {author} {\bibfnamefont {Akito}\ \bibnamefont {Sakai}}, \bibinfo
  {author} {\bibfnamefont {Satoru}\ \bibnamefont {Nakatsuji}}, \ and\ \bibinfo
  {author} {\bibfnamefont {Hisatomo}\ \bibnamefont {Harima}},\ }\bibfield
  {title} {\enquote {\bibinfo {title} {{Magnetic excitations and c-f
  hybridization effect in PrTi$_2$Al$_{20}$ and PrV$_2$Al$_{20}$}},}\
  }\href@noop {} {\bibfield  {journal} {\bibinfo  {journal} {Physical Review
  B}\ }\textbf {\bibinfo {volume} {88}},\ \bibinfo {pages} {085124} (\bibinfo
  {year} {2013})}\BibitemShut {NoStop}%
\bibitem [{\citenamefont {Shimura}\ \emph {et~al.}(2013)\citenamefont
  {Shimura}, \citenamefont {Ohta}, \citenamefont {Sakakibara}, \citenamefont
  {Sakai},\ and\ \citenamefont {Nakatsuji}}]{shimura2013evidence}%
  \BibitemOpen
  \bibfield  {author} {\bibinfo {author} {\bibfnamefont {Yasuyuki}\
  \bibnamefont {Shimura}}, \bibinfo {author} {\bibfnamefont {Yasuo}\
  \bibnamefont {Ohta}}, \bibinfo {author} {\bibfnamefont {Toshiro}\
  \bibnamefont {Sakakibara}}, \bibinfo {author} {\bibfnamefont {Akito}\
  \bibnamefont {Sakai}}, \ and\ \bibinfo {author} {\bibfnamefont {Satoru}\
  \bibnamefont {Nakatsuji}},\ }\bibfield  {title} {\enquote {\bibinfo {title}
  {{Evidence of a High-Field Phase in PrV$_2$Al${20}$ in a [100] Magnetic
  Field}},}\ }\href@noop {} {\bibfield  {journal} {\bibinfo  {journal} {J.
  Phys. Soc. Jpn.}\ }\textbf {\bibinfo {volume} {82}},\ \bibinfo {pages}
  {043705} (\bibinfo {year} {2013})}\BibitemShut {NoStop}%
\bibitem [{\citenamefont {Taniguchi}\ \emph {et~al.}(2016)\citenamefont
  {Taniguchi}, \citenamefont {Yoshida}, \citenamefont {Takeda}, \citenamefont
  {Takigawa}, \citenamefont {Tsujimoto}, \citenamefont {Sakai}, \citenamefont
  {Matsumoto},\ and\ \citenamefont {Nakatsuji}}]{taniguchi2016nmr}%
  \BibitemOpen
  \bibfield  {author} {\bibinfo {author} {\bibfnamefont {Takanori}\
  \bibnamefont {Taniguchi}}, \bibinfo {author} {\bibfnamefont {Makoto}\
  \bibnamefont {Yoshida}}, \bibinfo {author} {\bibfnamefont {Hikaru}\
  \bibnamefont {Takeda}}, \bibinfo {author} {\bibfnamefont {Masashi}\
  \bibnamefont {Takigawa}}, \bibinfo {author} {\bibfnamefont {Masaki}\
  \bibnamefont {Tsujimoto}}, \bibinfo {author} {\bibfnamefont {Akito}\
  \bibnamefont {Sakai}}, \bibinfo {author} {\bibfnamefont {Yosuke}\
  \bibnamefont {Matsumoto}}, \ and\ \bibinfo {author} {\bibfnamefont {Satoru}\
  \bibnamefont {Nakatsuji}},\ }\bibfield  {title} {\enquote {\bibinfo {title}
  {{NMR Observation of Ferro-Quadrupole Order in PrTi$_2$Al$_{20}$}},}\
  }\href@noop {} {\bibfield  {journal} {\bibinfo  {journal} {J. Phys. Soc.
  Jpn.}\ }\textbf {\bibinfo {volume} {85}},\ \bibinfo {pages} {113703}
  (\bibinfo {year} {2016})}\BibitemShut {NoStop}%
\bibitem [{\citenamefont {Ishii}\ \emph {et~al.}(2013)\citenamefont {Ishii},
  \citenamefont {Muneshige}, \citenamefont {Kamikawa}, \citenamefont {Fujita},
  \citenamefont {Onimaru}, \citenamefont {Nagasawa}, \citenamefont
  {Takabatake}, \citenamefont {Suzuki}, \citenamefont {Ano}, \citenamefont
  {Akatsu} \emph {et~al.}}]{ishii2013antiferroquadrupolar}%
  \BibitemOpen
  \bibfield  {author} {\bibinfo {author} {\bibfnamefont {Isao}\ \bibnamefont
  {Ishii}}, \bibinfo {author} {\bibfnamefont {Hitoshi}\ \bibnamefont
  {Muneshige}}, \bibinfo {author} {\bibfnamefont {Shuhei}\ \bibnamefont
  {Kamikawa}}, \bibinfo {author} {\bibfnamefont {Takahiro~K}\ \bibnamefont
  {Fujita}}, \bibinfo {author} {\bibfnamefont {Takahiro}\ \bibnamefont
  {Onimaru}}, \bibinfo {author} {\bibfnamefont {Naohiro}\ \bibnamefont
  {Nagasawa}}, \bibinfo {author} {\bibfnamefont {Toshiro}\ \bibnamefont
  {Takabatake}}, \bibinfo {author} {\bibfnamefont {Takashi}\ \bibnamefont
  {Suzuki}}, \bibinfo {author} {\bibfnamefont {Genki}\ \bibnamefont {Ano}},
  \bibinfo {author} {\bibfnamefont {Mitsuhiro}\ \bibnamefont {Akatsu}},  \emph
  {et~al.},\ }\bibfield  {title} {\enquote {\bibinfo {title}
  {{Antiferroquadrupolar ordering and magnetic-field-induced phase transition
  in the cage compound PrRh$_2$Zn$_{20}$}},}\ }\href@noop {} {\bibfield
  {journal} {\bibinfo  {journal} {Physical Review B}\ }\textbf {\bibinfo
  {volume} {87}},\ \bibinfo {pages} {205106} (\bibinfo {year}
  {2013})}\BibitemShut {NoStop}%
\bibitem [{\citenamefont {Higashinaka}\ \emph {et~al.}(2017)\citenamefont
  {Higashinaka}, \citenamefont {Nakama}, \citenamefont {Miyazaki},
  \citenamefont {Yamaura}, \citenamefont {Sato},\ and\ \citenamefont
  {Aoki}}]{higashinaka2017antiferroquadrupolar}%
  \BibitemOpen
  \bibfield  {author} {\bibinfo {author} {\bibfnamefont {Ryuji}\ \bibnamefont
  {Higashinaka}}, \bibinfo {author} {\bibfnamefont {Akihiro}\ \bibnamefont
  {Nakama}}, \bibinfo {author} {\bibfnamefont {Ryoichi}\ \bibnamefont
  {Miyazaki}}, \bibinfo {author} {\bibfnamefont {Jun-ichi}\ \bibnamefont
  {Yamaura}}, \bibinfo {author} {\bibfnamefont {Hideyuki}\ \bibnamefont
  {Sato}}, \ and\ \bibinfo {author} {\bibfnamefont {Yuji}\ \bibnamefont
  {Aoki}},\ }\bibfield  {title} {\enquote {\bibinfo {title}
  {{Antiferroquadrupolar Ordering in Quadrupolar Kondo Lattice of Non-Kramers
  System PrTa2Al20}},}\ }\href@noop {} {\bibfield  {journal} {\bibinfo
  {journal} {Journal of the Physical Society of Japan}\ }\textbf {\bibinfo
  {volume} {86}},\ \bibinfo {pages} {103703} (\bibinfo {year}
  {2017})}\BibitemShut {NoStop}%
\bibitem [{Note1()}]{Note1}%
  \BibitemOpen
  \bibinfo {note} {Akito Sakai, Talk given at the J-Physics Topical Meeting on
  ``Exotic Phenomena in Itinerant Multipole Systems'', ISSP, University of
  Tokyo, December 18, 2017}\BibitemShut {NoStop}%
\bibitem [{\citenamefont {Hattori}\ and\ \citenamefont
  {Tsunetsugu}(2014)}]{hattori2014antiferro}%
  \BibitemOpen
  \bibfield  {author} {\bibinfo {author} {\bibfnamefont {Kazumasa}\
  \bibnamefont {Hattori}}\ and\ \bibinfo {author} {\bibfnamefont {Hirokazu}\
  \bibnamefont {Tsunetsugu}},\ }\bibfield  {title} {\enquote {\bibinfo {title}
  {{Antiferro Quadrupole Orders in Non-Kramers Doublet Systems}},}\ }\href@noop
  {} {\bibfield  {journal} {\bibinfo  {journal} {J. Phys. Soc. Jpn.}\ }\textbf
  {\bibinfo {volume} {83}},\ \bibinfo {pages} {034709} (\bibinfo {year}
  {2014})}\BibitemShut {NoStop}%
\bibitem [{\citenamefont {Hattori}\ and\ \citenamefont
  {Tsunetsugu}(2016)}]{hattori2016antiferro}%
  \BibitemOpen
  \bibfield  {author} {\bibinfo {author} {\bibfnamefont {Kazumasa}\
  \bibnamefont {Hattori}}\ and\ \bibinfo {author} {\bibfnamefont {Hirokazu}\
  \bibnamefont {Tsunetsugu}},\ }\bibfield  {title} {\enquote {\bibinfo {title}
  {{Classical Monte Carlo Study for Antiferro Quadrupole Orders in a Diamond
  Lattice}},}\ }\href@noop {} {\bibfield  {journal} {\bibinfo  {journal} {J.
  Phys. Soc. Jpn.}\ }\textbf {\bibinfo {volume} {85}},\ \bibinfo {pages}
  {094001} (\bibinfo {year} {2016})}\BibitemShut {NoStop}%
\bibitem [{\citenamefont {Freyer}\ \emph {et~al.}(2018)\citenamefont {Freyer},
  \citenamefont {Attig}, \citenamefont {Lee}, \citenamefont {Paramekanti},
  \citenamefont {Trebst},\ and\ \citenamefont {Kim}}]{freyer2018two}%
  \BibitemOpen
  \bibfield  {author} {\bibinfo {author} {\bibfnamefont {Frederic}\
  \bibnamefont {Freyer}}, \bibinfo {author} {\bibfnamefont {Jan}\ \bibnamefont
  {Attig}}, \bibinfo {author} {\bibfnamefont {SungBin}\ \bibnamefont {Lee}},
  \bibinfo {author} {\bibfnamefont {Arun}\ \bibnamefont {Paramekanti}},
  \bibinfo {author} {\bibfnamefont {Simon}\ \bibnamefont {Trebst}}, \ and\
  \bibinfo {author} {\bibfnamefont {Yong~Baek}\ \bibnamefont {Kim}},\
  }\bibfield  {title} {\enquote {\bibinfo {title} {{Two-stage multipolar
  ordering in PrT$_2$Al$_{20}$ Kondo materials}},}\ }\href@noop {} {\bibfield
  {journal} {\bibinfo  {journal} {Physical Review B}\ }\textbf {\bibinfo
  {volume} {97}},\ \bibinfo {pages} {115111} (\bibinfo {year}
  {2018})}\BibitemShut {NoStop}%
\bibitem [{\citenamefont {Stevens}(1952)}]{stevens1952matrix}%
  \BibitemOpen
  \bibfield  {author} {\bibinfo {author} {\bibfnamefont {K.W.H.}\ \bibnamefont
  {Stevens}},\ }\bibfield  {title} {\enquote {\bibinfo {title} {Matrix elements
  and operator equivalents connected with the magnetic properties of rare earth
  ions},}\ }\href@noop {} {\bibfield  {journal} {\bibinfo  {journal}
  {Proceedings of the Physical Society. Section A}\ }\textbf {\bibinfo {volume}
  {65}},\ \bibinfo {pages} {209} (\bibinfo {year} {1952})}\BibitemShut
  {NoStop}%
\bibitem [{\citenamefont {Lea}\ \emph {et~al.}(1962)\citenamefont {Lea},
  \citenamefont {Leask},\ and\ \citenamefont {Wolf}}]{lea1962raising}%
  \BibitemOpen
  \bibfield  {author} {\bibinfo {author} {\bibfnamefont {K.R.}\ \bibnamefont
  {Lea}}, \bibinfo {author} {\bibfnamefont {M.J.M.}\ \bibnamefont {Leask}}, \
  and\ \bibinfo {author} {\bibfnamefont {W.P.}\ \bibnamefont {Wolf}},\
  }\bibfield  {title} {\enquote {\bibinfo {title} {{The raising of angular
  momentum degeneracy of f-electron terms by cubic crystal fields}},}\
  }\href@noop {} {\bibfield  {journal} {\bibinfo  {journal} {Journal of Physics
  and Chemistry of Solids}\ }\textbf {\bibinfo {volume} {23}},\ \bibinfo
  {pages} {1381} (\bibinfo {year} {1962})}\BibitemShut {NoStop}%
\bibitem [{\citenamefont {Wu}(1982)}]{wu1982potts}%
  \BibitemOpen
  \bibfield  {author} {\bibinfo {author} {\bibfnamefont {Fa-Yueh}\ \bibnamefont
  {Wu}},\ }\bibfield  {title} {\enquote {\bibinfo {title} {{The Potts
  model}},}\ }\href@noop {} {\bibfield  {journal} {\bibinfo  {journal} {Reviews
  of Modern Physics}\ }\textbf {\bibinfo {volume} {54}},\ \bibinfo {pages}
  {235} (\bibinfo {year} {1982})}\BibitemShut {NoStop}%
\bibitem [{\citenamefont {Hove}\ and\ \citenamefont
  {Sudb{\o}}(2003)}]{hove2003criticality}%
  \BibitemOpen
  \bibfield  {author} {\bibinfo {author} {\bibfnamefont {J}~\bibnamefont
  {Hove}}\ and\ \bibinfo {author} {\bibfnamefont {A}~\bibnamefont {Sudb{\o}}},\
  }\bibfield  {title} {\enquote {\bibinfo {title} {{Criticality versus q in the
  (2+1)-dimensional Z$_q$ clock model}},}\ }\href@noop {} {\bibfield  {journal}
  {\bibinfo  {journal} {Physical Review E}\ }\textbf {\bibinfo {volume} {68}},\
  \bibinfo {pages} {046107} (\bibinfo {year} {2003})}\BibitemShut {NoStop}%
\bibitem [{\citenamefont {Bellafard}\ \emph {et~al.}(2015)\citenamefont
  {Bellafard}, \citenamefont {Chakravarty}, \citenamefont {Troyer},\ and\
  \citenamefont {Katzgraber}}]{Bellafard2015}%
  \BibitemOpen
  \bibfield  {author} {\bibinfo {author} {\bibfnamefont {Arash}\ \bibnamefont
  {Bellafard}}, \bibinfo {author} {\bibfnamefont {Sudip}\ \bibnamefont
  {Chakravarty}}, \bibinfo {author} {\bibfnamefont {Matthias}\ \bibnamefont
  {Troyer}}, \ and\ \bibinfo {author} {\bibfnamefont {Helmut~G.}\ \bibnamefont
  {Katzgraber}},\ }\bibfield  {title} {\enquote {\bibinfo {title} {{The effect
  of quenched bond disorder on first-order phase transitions}},}\ }\href@noop
  {} {\bibfield  {journal} {\bibinfo  {journal} {Annals of Physics}\ }\textbf
  {\bibinfo {volume} {357}},\ \bibinfo {pages} {66 -- 78} (\bibinfo {year}
  {2015})}\BibitemShut {NoStop}%
\end{thebibliography}

%


\appendix

\end{document}